\documentclass[submission, Phys]{SciPost}
\usepackage{amsmath,amssymb,mathtools,xspace, mathrsfs}
\usepackage{booktabs,multirow,graphicx,tabularx,slashed}
\usepackage{hyperref}
\usepackage{color,xcolor}
\usepackage[normalem]{ulem}
\usepackage{enumitem}
\usepackage{braket}
\usepackage{stackrel}
\usepackage{tikz}
\usepackage{float}
\usepackage{bm}

\graphicspath{{plots/}}

\makeatletter
\def\BState{\State\hskip-\ALG@thistlm}
\makeatother

\makeatletter
\@ifundefined{pdfoutput}{}{\DeclareGraphicsRule{*}{mps}{*}{}}
\makeatother

\makeatletter
\DeclareRobustCommand*{\bfseries}{%
   \not@math@alphabet\bfseries\mathbf
   \fontseries\bfdefault\selectfont
   \boldmath
}
\makeatother

\setitemize{itemsep=2pt,topsep=2pt,parsep=0pt,partopsep=0pt,leftmargin=*}
\setenumerate{itemsep=0pt,topsep=2pt,parsep=0pt,partopsep=0pt,labelindent=3pt,leftmargin=*}
\definecolor{Gcolor}{HTML}{3b528b}
\definecolor{Dcolor}{HTML}{e41a1c}

\tikzstyle{generator} = [rectangle, rounded corners, minimum width=3cm, minimum height=1cm,text centered, draw=Gcolor]
\tikzstyle{discriminator} = [rectangle, rounded corners, minimum width=3cm, minimum height=1cm,text centered, draw=Dcolor]
\tikzstyle{io} = [circle, trapezium left angle=70, trapezium right angle=110, minimum width=1cm, minimum height=1cm, text centered, draw=black]

\tikzstyle{process} = [rectangle, minimum width=1cm, minimum height=1cm, text centered, draw=black]
\tikzstyle{decision} = [rectangle, minimum width=1cm, minimum height=1cm, text centered, draw=black]

\tikzstyle{arrow} = [thick,->,>=stealth]
\usepackage{xcolor}

\newcommand{\XLangle}{\Big\langle}
\newcommand{\XRangle}{\Big\rangle}

\newcommand\one{\leavevmode\hbox{\small1\normalsize\kern-.33em1}}

\newcommand{\loss}{\mathcal{L}}

\newcommand{\qqquad}{\qquad \qquad}

\newcommand{\gev}{\text{GeV}}

\def\slashchar#1{\setbox0=\hbox{$#1$}           % set a box for #1
   \dimen0=\wd0                                 % and get its size
   \setbox1=\hbox{/} \dimen1=\wd1               % get size of /
   \ifdim\dimen0>\dimen1                        % #1 is bigger
      \rlap{\hbox to \dimen0{\hfil/\hfil}}      % so center / in box
      #1                                        % and print #1
   \else                                        % / is bigger
      \rlap{\hbox to \dimen1{\hfil$#1$\hfil}}   % so center #1
      /                                         % and print /
   \fi}

\begin{document}
%\begin{fmffile}{feynman}

\begin{center}{\Large \textbf{
      Generative Networks for Precision Enthusiasts
}}\end{center}

\begin{center}
Anja Butter\textsuperscript{1}, 
Theo Heimel\textsuperscript{1}, 
Sander Hummerich\textsuperscript{1}, 
Tobias Krebs\textsuperscript{1}, \\
Tilman Plehn\textsuperscript{1}, 
Armand Rousselot\textsuperscript{2}, and
Sophia Vent\textsuperscript{1}
\end{center}

\begin{center}
{\bf 1} Institut f\"ur Theoretische Physik, Universit\"at Heidelberg, Germany \\
{\bf 2} Heidelberg Collaboratory for Image Processing, Universit\"at Heidelberg, Germany
\end{center}

\begin{center}
\today
\end{center}

\section*{Abstract}
         {\bf Generative networks are opening new avenues in fast
           event generation for the LHC. We show how generative flow
           networks can reach percent-level precision for kinematic
           distributions, how they can be trained jointly with a
           discriminator, and how this discriminator improves the
           generation. Our joint training relies on a novel coupling
           of the two networks which does not require a Nash
           equilibrium. We then estimate the generation uncertainties
           through a Bayesian network setup and through conditional
           data augmentation, while the discriminator ensures that
           there are no systematic inconsistencies compared to the
           training data.}

\vspace{10pt}
\noindent\rule{\textwidth}{1pt}
\tableofcontents\thispagestyle{fancy}
\noindent\rule{\textwidth}{1pt}
\vspace{10pt}

\newpage
%%%%%%%%%%%%%%%%%%%%%%%%%%%%%%%%%%%%%%%%%%%%%%%%%%%%%%%%%%%%%%%%%%%%%%%%
\section{Introduction}
\label{sec:intro}

Precise first-principle simulations provided by the theory community
are a defining feature of Large Hadron Collider (LHC) physics. They
are based on perturbative quantum field theory with fundamental
Lagrangians as their physics input, and they provide the simulated
events necessary for modern LHC analyses. Because of the close
correlation of complexity and precision in perturbative calculations,
precision and speed are, largely, two sides of the same medal. Both of
these sides are facing major challenges for the LHC Runs~3 and~4, and
the hope is that machine learning and its modern numerics toolbox
allow us to provide the simulations needed for a 25-fold increase of
LHC data as compared to Run~2.

In recent years, modern machine learning has shown great potential to
improve LHC simulations. Underlying techniques include generative
adversarial networks
(GANs)~\cite{goodfellow2014generative,Creswell2018,Butter:2020qhk},
variational autoencoders
(VAEs)~\cite{kingma2014autoencoding,Kingma2019}, normalizing
flows~\cite{nflow1,Kobyzev_2020,papamakarios2019normalizing,nflow_review,mller2018neural},
and their invertible network (INN)
variant~\cite{inn,coupling2,glow}. As part of the standard LHC event
generation chain~\cite{Butter:2020tvl}, modern neural networks can be
applied to the full range of phase space
integration~\cite{maxim,Chen:2020nfb}, phase space
sampling~\cite{Bothmann:2020ywa,Gao:2020vdv,Gao:2020zvv,Danziger:2021eeg},
amplitude computations~\cite{Bishara:2019iwh,Badger:2020uow}, event
subtraction~\cite{Butter:2019eyo}, event
unweighting~\cite{Verheyen:2020bjw,Backes:2020vka}, parton
showering~\cite{shower,locationGAN,monkshower,juniprshower,Dohi:2020eda},
or super-resolution enhancement~\cite{DiBello:2020bas,Baldi:2020hjm}.
Conceptionally new developments are, for instance, based on fully
NN-based event
generators~\cite{dutch,gan_datasets,DijetGAN2,Butter:2019cae,Alanazi:2020klf}
or detector
simulations~\cite{calogan1,calogan2,fast_accurate,aachen_wgan1,aachen_wgan2,ATLASsimGAN,Belayneh:2019vyx,Buhmann:2020pmy,Buhmann:2021lxj,Chen:2021gdz,Krause:2021ilc}.
In essence, there is no aspect of the standard event generation chain
that cannot be improved through modern machine learning.

A structural advantage of generative networks for event generation or
detector simulations is that, unlike forward Monte Carlo simulations,
the network-based generation can be inverted. Specifically,
conditional GANs and INNs allow us to invert the simulation chain to
unfold detector effects~\cite{Datta:2018mwd,fcgan} and extract the
hard scattering process at parton level in a statistically consistent
manner~\cite{Bellagente:2020piv}. Because of their superior
statistical properties, the same conditional INNs can be used for
simulation-based inference based on high-dimensional and low-level
data~\cite{Bieringer:2020tnw}.  Finally, normalizing-flow or INN
generators provide new opportunities when we combine them with
Bayesian network
concepts\cite{bnn_early,bnn_early2,bnn_early3,deep_errors,Bollweg:2019skg,Kasieczka:2020vlh}
to construct uncertainty-controlled generative
networks~\cite{Bellagente:2021yyh}.

In this paper we combine the full range of ML-concepts to build an
NN-based LHC event generator which meets the requirements in terms of
phase space coverage, precision, and control of different
uncertainties.  We first present a precision INN generator in
Sec.~\ref{sec:prec} which learns underlying phase space densities such
that kinematic distributions are reproduced at the percent level,
consistent with the statistical limitations of the training
data. Next, our inspiration by GANs leads us to construct the DiscFlow
discriminator--generator architecture to control the consistency of
training data and generative network in in
Sec.~\ref{sec:gan}. Finally, in Sec.~\ref{sec:uncert} we illustrate
three ways to control the network training and estimate remaining
uncertainties (i) through a Bayesian generative network, (ii) using
conditional augmentations for systematic or theory uncertainties, and
(iii) using the DiscFlow discriminator for controlled reweighting.
While we employ forward event generation to illustrate these different
concepts, our results can be directly transferred to inverted
simulation, unfolding, or inference problems.

%%%%%%%%%%%%%%%%%%%%%%%%%%%%%%%%%%%%%%%%%%%%%%%%%%%%%%%%%%%%%%%%%%%%%%%%
\section{Precision generator}
\label{sec:prec}

As we will show in this paper, generative networks using normalizing
flows have significant advantages over other network architectures,
including GANs, when it comes to LHC event generation. As a starting
point, we show how flow-based invertible networks can be trained to
generate events and reproduce phase space densities with high
precision. Our network architecture accounts for the complication of a
variable number of particles in the final state.

%%%%%%%%%%%%%%%%%%%%%%%%%%%%%%%%%%%%%%%%%%%%%%%%%%%%%%%%%%%%%%%%%%%%%%%%
\subsection{Data set}
\label{sec:prec_data}

The kind of NN-generators we discuss in this paper are trained on
unweighted events at the hadronization level. We exclude detector
effects because they soften sharp phase space features, so simulations
without them tend to be more challenging and their results are more
interesting from a technical perspective. This means our method will
work even better on reconstucted objects.

The production of leptonically decaying $Z$-bosons with a variable
number of jets is an especially challenging benchmark process. First,
the network has to learn an extremely sharp $Z$-resonance
peak. Second, QCD forces us to apply a geometric separation between
jets, inducing a non-trivial topology of phase space. Finally, again
because of QCD it does not make sense to define final states with a
fixed number of jets, so our generative network has to cover a final
state with a variable number of dimensions. Given these considerations
we work with the process
\begin{align}
pp \to Z_{\mu \mu} + \{ 1,2,3 \}~\text{jets} \; ,
\label{eq:ref_proc}
\end{align}
simulated with \textsc{Sherpa}2.2.10~\cite{Sherpa:2019gpd} at
13~TeV. We use CKKW merging~\cite{Catani:2001cc} to generate a merged
sample with up to three hard jets including ISR, parton shower,
and hadronization, but no pile-up. The final state of the training
sample is defined by \textsc{Fastjet}3.3.4~\cite{fastjet} in terms of
anti-$k_T$ jets~\cite{anti-kt} with
\begin{align}
  p_{T,j} > 20~\gev
  \qqquad \text{and} \qqquad
  \Delta R_{jj} > R_\text{min} = 0.4 \; .
\end{align}
The jets and muons are ordered in $p_T$. Because jets have a finite
invariant mass, our final state dimensionality is three for each muon
plus four degrees of freedom per jet, giving us phase space
dimensionalities 10, 14, and 18.  Momentum conservation does not
further reduce the dimensionality, as not every generated hadron is
captured by the three leading jets. However, we will reduce this
dimensionality by one by removing the symmetry on the choice of global
azimuthal angle.  Our combined sample size is 5.4M events, divided
into 4.0M one-jet events, 1.1M two-jet events, and 300k three-jet
events. This different training statistics will be discussed in more
detail in Sec.~\ref{sec:uncert_bayes}.\bigskip

To define a representation which makes it easier for an INN to learn
the kinematic patterns we apply a standard pre-processing.
First, each lepton or reconstructed jet is represented by
\begin{align}
    \{ \; p_T, \eta, \phi, m \; \} \; .
\label{eq:def_obs}
\end{align}
Because we can extract a global threshold in the jet $p_T$ we
represent the events in terms of the variable $\tilde{p}_T = \log (
p_T - p_{T,\text{min}})$. This form leads to an approximately Gaussian
distribution, matching the Gaussian latent-space distribution of the
INN. Second, the choice of the global azimuthal angle is a symmetry of
LHC events, so we instead train on azimuthal angles relative to the
muon with larger transverse momentum in the range $\Delta \phi \in [-\pi, \pi]$. A
transformation into $\widetilde{\Delta\phi} = \text{atanh} (
\Delta\phi/\pi )$ again leads to an approximately Gaussian
distribution.  For all phase space variables $q$ we apply a
centralization and normalization step
\begin{align}
  \tilde{q}_i = \frac{q_i - \overline{q_i}}{\sigma(q_i)} \; .
\end{align}
Finally, we apply a whitening/PCA transformation separately for each
jet multiplicity.

%%%%%%%%%%%%%%%%%%%%%%%%%%%%%%%%%%%%%%%%%%%%%%%%%%%%%%%%%%%%%%%%%%%%%%%%
\subsection{INN generator}
\label{sec:prec_arch}

%----------------------------------------------------------
\begin{figure}[t]
  \centering
  \usetikzlibrary{shapes.geometric, arrows, arrows.meta,shapes}

\definecolor{Bcolor}{HTML}{0081FA}
\definecolor{Ycolor}{HTML}{FFE699}

\tikzstyle{network} = [thick, rectangle, minimum width=2cm, minimum height=1.5cm, text centered, align=center, fill=Ycolor, draw]
\tikzstyle{conn} = [ellipse, thick, inner sep=0, Bcolor, draw]
\tikzstyle{arrow} = [thick, Bcolor, -{Latex[scale=1.0]}] %, line width=0.5mm]

\begin{tikzpicture}[node distance=2cm] %, scale=0.55, every node/.style={transform shape}]

\node (cinn1) [network] {cINN};
\node (cinn2) [network, below of=cinn1, xshift=2.5cm] {cINN};
\node (cinn3) [network, below of=cinn2, xshift=2.5cm] {cINN};

\draw [color=black] ([xshift=-1cm, yshift=1.5cm]cinn1.west) -- ([xshift=6cm, yshift=1.5cm]cinn1.east);
\draw [color=black] ([xshift=-1cm, yshift=1.7cm]cinn1.west) -- ([xshift=6cm, yshift=1.7cm]cinn1.east);
\draw [color=black] ([xshift=-1cm, yshift=1.3cm]cinn1.west) -- ([xshift=6cm, yshift=1.3cm]cinn1.east);

\draw [color=black] ([xshift=-1cm]cinn1.west) -- ([yshift=0cm]cinn1.west);
\draw [color=black] ([xshift=-1cm, yshift=-0.2cm]cinn1.west) -- ([yshift=-0.2cm]cinn1.west);
\draw [color=black] ([xshift=-1cm, yshift=0.2cm]cinn1.west) -- ([yshift=0.2cm]cinn1.west);
\draw [color=black] ([xshift=6cm]cinn1.east) -- ([yshift=0cm]cinn1.east);
\draw [color=black] ([xshift=6cm, yshift=-0.2cm]cinn1.east) -- ([yshift=-0.2cm]cinn1.east);
\draw [color=black] ([xshift=6cm, yshift=0.2cm]cinn1.east) -- ([yshift=0.2cm]cinn1.east);

\draw [color=black] ([xshift=-3.5cm]cinn2.west) -- ([yshift=0cm]cinn2.west);
\draw [color=black] ([xshift=3.5cm]cinn2.east) -- ([yshift=0cm]cinn2.east);

\draw [color=black] ([xshift=-6cm]cinn3.west) -- ([yshift=0cm]cinn3.west);
\draw [color=black] ([xshift=1cm]cinn3.east) -- ([yshift=0cm]cinn3.east);

\node [left of=cinn1, xshift=-0.6cm, yshift=1.5cm] {$N_\mathrm{jets}$};
\node [left of=cinn1, xshift=-0.6cm] {$z_{1 \ldots 9}$};
\node [left of=cinn2, xshift=-3.1cm] {$z_{10 \ldots 13}$};
\node [left of=cinn3, xshift=-5.6cm] {$z_{14 \ldots 17}$};

\node [right of=cinn1, xshift=5.8cm, yshift=1.5cm] {(1-hot)};
\node [right of=cinn1, xshift=6.0cm] {$\mu_1,\mu_2,j_1$};
\node [right of=cinn2, xshift=2.9cm] {$j_2$};
\node [right of=cinn3, xshift=0.4cm] {$j_3$};

\node (c1) [conn, above of=cinn1, minimum width=0.15cm, minimum height=0.6cm, yshift=-0.5cm] {};
\node (c2) [conn, above of=cinn2, minimum width=0.15cm, minimum height=0.4cm, yshift=1.4cm, xshift=-0.15cm] {};
\node (c3) [conn, above of=cinn2, minimum width=0.15cm, minimum height=0.6cm, yshift=0.0cm, xshift=0.15cm] {};
\node (c4) [conn, above of=cinn3, minimum width=0.15cm, minimum height=0.6cm, yshift=2.0cm, xshift=-0.15cm] {};
\node (c5) [conn, above of=cinn3, minimum width=0.15cm, minimum height=0.15cm, yshift=0.0cm, xshift=0.15cm] {};

\draw [arrow] ([xshift=0.0cm]c1.south) -- ([xshift=0.0cm]cinn1.north);
\draw [arrow] ([xshift=0.0cm]c2.south) -- ([xshift=-0.15cm]cinn2.north);
\draw [arrow] ([xshift=0.0cm]c3.south) -- ([xshift=0.15cm]cinn2.north);
\draw [arrow] ([xshift=0.0cm]c4.south) -- ([xshift=-0.15cm]cinn3.north);
\draw [arrow] ([xshift=0.0cm]c5.south) -- ([xshift=0.15cm]cinn3.north);

\end{tikzpicture}
  \caption{Generative flow architecture for events with two muons and
    one to three jets. The INNs relate the latent space (left) to the
    physical phase space (right).}
  \label{fig:architecture}
\end{figure}
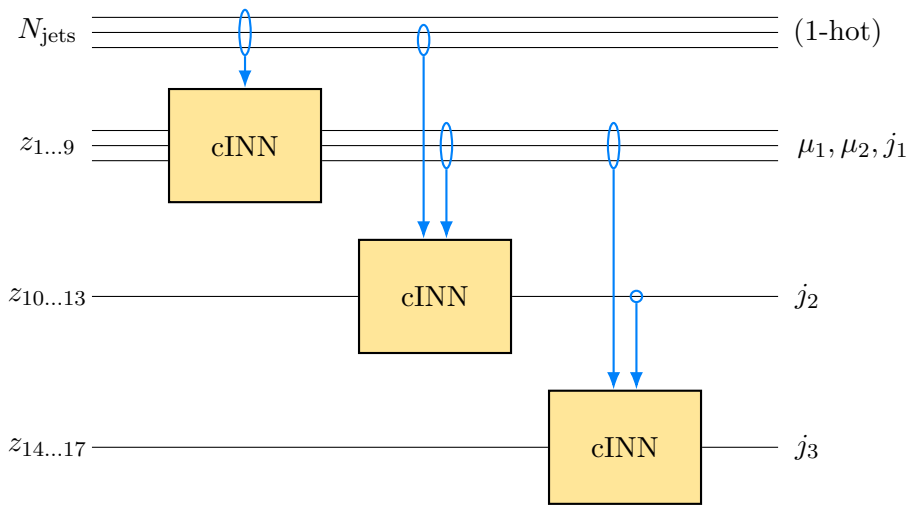
%----------------------------------------------------------

%----------------------------------------------------------
\begin{table}[b!]
\centering
\begin{small}
\begin{tabular}{l|cccc}
\toprule
    hyperparameter 		& INN (Sec.~\ref{sec:prec_arch}) & INN (Sec.~\ref{sec:gan_disc}) & DiscFlow (Sec.~\ref{sec:gan_joint}) & BINN  (Sec.~\ref{sec:uncert_bayes}) \\
    \midrule
    LR scheduling                      & one-cycle & same & same & same \\
    Starter LR			       & $10^{-4}$	& $4\cdot10^{-4}$ & $2\cdot10^{-4}$ & $10^{-5}$\\
    Maximum LR                         & $10^{-3}$	& $4\cdot10^{-3}$& $2\cdot10^{-3}$ & $10^{-4}$\\\
    Epochs                             & 100  & 200 & 200 & 100\\
    Batch size			       & 1024	& 2048 & 2048 & 3072\\
    \textsc{Adam} $\beta_1$, $\beta_2$ & $0.9$, $0.99$ & $0.9$, $0.99$ & $0.5$, $0.9$ & same \\ \midrule
    Coupling block                     & cubic spline & same & same & same \\
    \# spline bins                     & 60 & same & same & same \\
    \# coupling blocks                 & 25 & 25 & 25 & 20 \\
    Layers per block                   & 3  & 3 & 3 & 6 \\ \midrule
    \# generated events                & 2M& 2M & 2M & 1M \\
    \bottomrule
\end{tabular}
\end{small}
\caption{Training setup and hyperparameters for the INN generators
  used in our different setups.}
\label{tab:setup}
\end{table}
%----------------------------------------------------------

For a fixed final-state dimensionality we can use a completely
standard INN~\cite{inn,Bellagente:2021yyh} to generate LHC events,
especially after the preprocessing step defined above. Our technical
challenge is to expand the INN architecture to generate final states
with 9, 13, and 17 phase space dimensions. Of course, we could just
split the training sample into different multiplicities and train a
set of individual networks. However, in this case each of these
networks has to learn the basic QCD patterns, making this naive
approach inefficient and unstable.

To increase the efficiency of the training, we use one network for the
common $\mu_{1,2}$ and $j_1$ momenta and add additional small networks
for each additional jet, as illustrated in
Fig.~\ref{fig:architecture}. Some basic kinematic features of the
muons and the first jet, like their transverse momentum balance,
depend on possible additional jets, so we first provide the base
network with the one-hot encoded number of jets as a condition. This
allows the base network to generate all relevant $\{ \mu \mu j
\}$-configurations. Starting from those configurations we then train
additional networks for each additional jet. These small networks are
conditioned on the training observables of the base networks or the
lower-multiplicity network, and on the number of jets. Because the
$\mu \mu j$ and $\mu \mu jj$ networks are trained on events with mixed
multiplicities, we guarantee a balanced training by drawing a random
subset of the training data set at the beginning of each epoch
containing equal numbers of events from all different multiplicities.
While all three networks are trained separately, they are combined as
a generator. We have found this conditional network architecture to
provide the best balance of training time and performance.

Our network is implemented using \textsc{PyTorch}~\cite{pytorch} with
the \textsc{Adam} optimizer~\cite{adam}, and a one-cycle learning-rate
scheduler~\cite{one-cycle-lr}. The affine coupling blocks of the standard conditional INN
setup~\cite{cinn,Bellagente:2020piv} are replaced by cubic
spline coupling blocks~\cite{durkan2019cubic}, which are more efficient in learning complex phase space
patterns precisely and reliably. The coupling block splits the target space into bins of variable width based on trainable support points, which are connected with a cubic function. They are combined with random but fixed rotations to ensure interaction between all input variables. The parameter ranges of input, output and intermediate
spaces are limited to $[-10,10]$ on both sides of the coupling blocks,
numbers outside this range are mapped onto themselves. The individual
coupling blocks split their input vector in two halves $(u_i, v_i)$ 
and transforms $v_i$ as
\begin{equation}
    {v_i}' = s(v_i; \chi(u_i, c_i)) \; .
\end{equation}
The $c_i$ are the conditional inputs of the network. The function
$\chi$ is a fully connected sub-network with $2 n_\text{bins} + 2$
outputs, where $n_\text{bins}$ is the number of spline bins. They
encode the horizontal and vertical positions of the spline knots and
its slope at the boundaries. The loss function for a cINN can most
easily be defined in terms of the ratio of the intractable reference
density $P_\text{data}(x;c)$ and the learned or model density $P(x;c)$
in terms of the phase space position $x$ and the condition $c$.  We
can ignore the normalization $\log P_\text{data}(x;c)$, because it
does not affect the network training,
\begin{align}
  \loss_G
  &= - \int dx \; P_\text{data}(x,c) \; \log \frac{P(x;c)}{P_\text{data}(x;c)} \notag \\
  &= - \int dx \; P_\text{data}(x,c) \; \log P(x;c) + \text{const} \notag \\
  &= - \int dx \; P_\text{data}(x,c) \; \Big[ \log P_\text{latent}(\psi(x;c)) + \log J(x;c) \Big] + \text{const} 
\label{eq:inn_loss_cont}
\end{align}
In the last line we change variables between phase space and latent
space and split $P(x;c)$ into an the latent-space distribution in
terms of the INN-encoded mapping $\psi$ and its Jacobian $J$. Assuming
a Gaussian in the latent space this gives us for a batch of $B$ inputs
\begin{align}
  \loss_G 
  \approx \sum_{i=1}^B \left( \frac{\psi(x_i; c_i)^2}{2}
    - \log J_i \right) \; .
    \label{eq:inn_loss}
\end{align}
We list all
hyperparameters in Tab.~\ref{tab:setup}.

%
%\begin{align}
%  \loss_G
%  &= - \sum_{i=1}^B \log \frac{P(x_i;c_i)}{P_\text{data}(x_i;c_i)} 
%  = - \sum_{i=1}^B \log P(x_i;c_i) + \text{const}
%  \; ,
%\label{eq:inn_loss_complete}
%\end{align}
%

%----------------------------------------------------------
\begin{figure}[t]
    \centering
    \includegraphics[page = 7,width=0.99\textwidth]{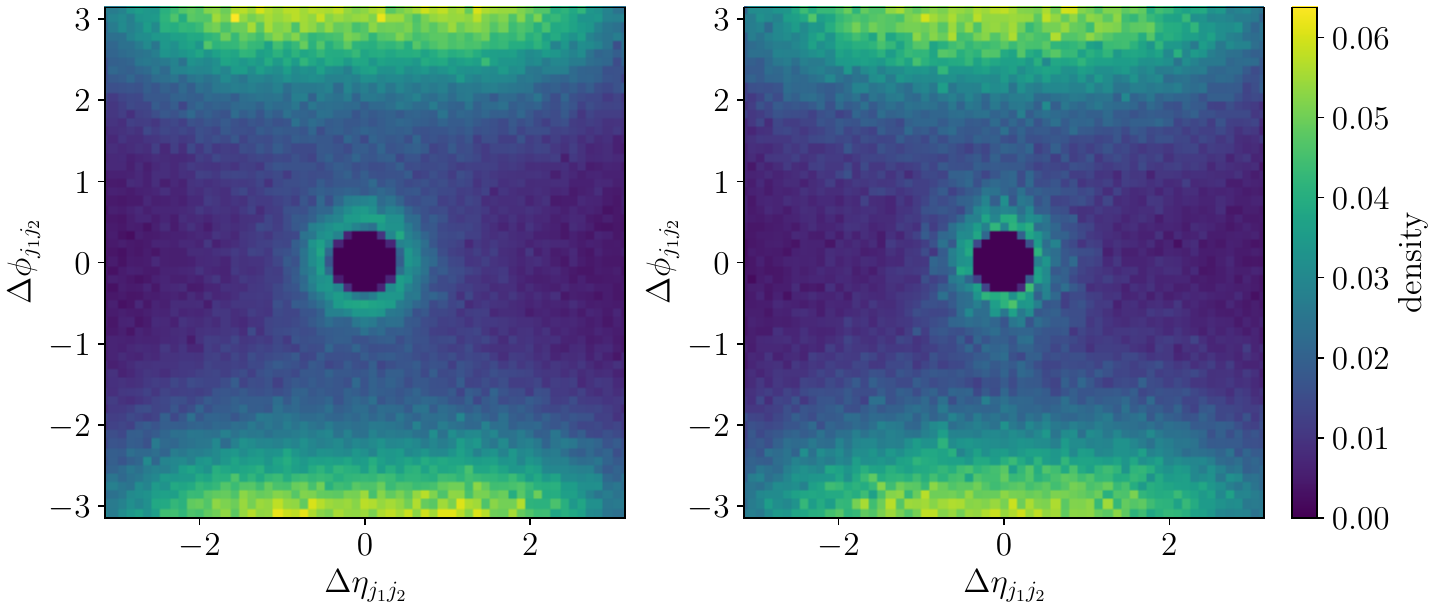}
    \caption{Jet-jet correlations for events with two jets. We show
      truth (left) and INN-generated events (right).} 
    \label{fig:inn_2d}
\end{figure}
%----------------------------------------------------------

%%%%%%%%%%%%%%%%%%%%%%%%%%%%%%%%%%%%%%%%%%%%%%%%%%%%%%%%%%%%%%%%%%%%%%%%
\subsubsection*{Magic transformation}

%----------------------------------------------------------
\begin{figure}[t]
    \includegraphics[width=0.495\textwidth]{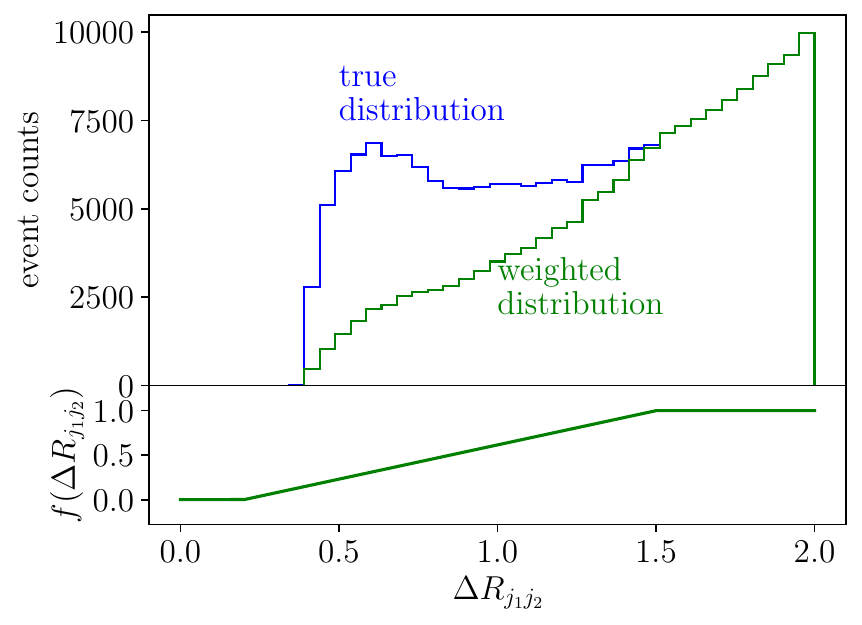}
    \includegraphics[width=0.495\textwidth]{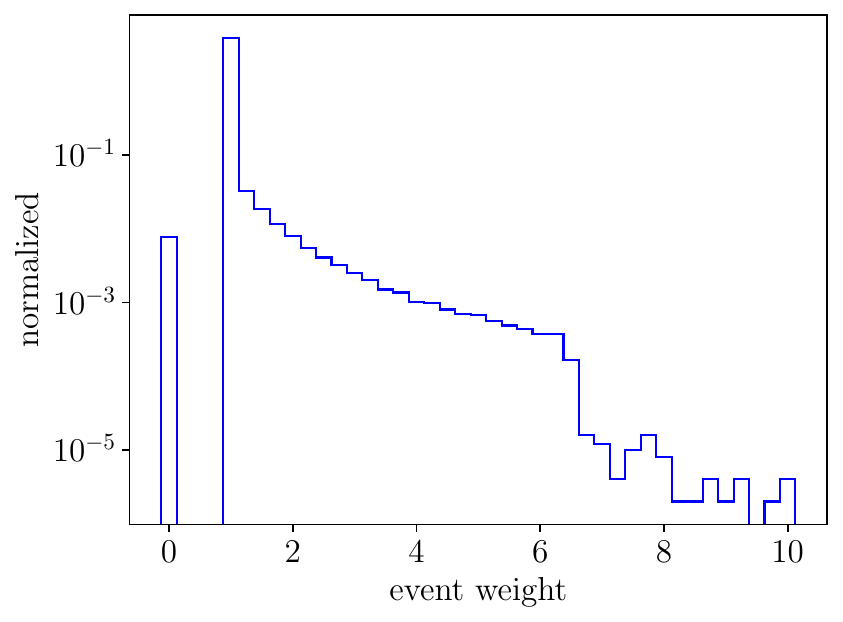}
    \caption{Left: $\Delta R_{j_1 j_2}$-distribution for $Z+2$~jets
      events before and after the transformation of
      Eq.\eqref{eq:trafo}. Right: histogram of the weights of the
      generated events.}
    \label{fig:trafo}
\end{figure}
%----------------------------------------------------------

A major challenge of the $Z+$~jets final state is illustrated in
Fig.~\ref{fig:inn_2d}, where we show the $\Delta \phi$ vs $\Delta
\eta$ correlations for the exclusive 2-jet sample. We see that most
events prefer a back-to-back topology, but a small number of events
features two jets recoiling against the $Z$, cut off by the
requirement $\Delta R_{jj} > 0.4$. The ring around the corresponding
small circle is a local maximum, and inside the ring the phase space
density drops to zero. Because this entire structure lives in a poorly
populated phase space region, the INN typically ignores the local
maximum and smoothly interpolates over the entire ring-hole
structure. We emphasize that in our case this problem is not caused by
the non-trivial phase space topology~\cite{Winterhalder:2021ave}, the
network interpolates smoothly through the holes, but a problem of the
precision with which the network learns features just around these
holes.

We can improve our network performance, after noticing the issue, by
using some physics intuition and observing a near-magic aspect of
network training. To this end, we map out the local maximum structure
and make use of the fact that our network is extremely efficient at
interpolating smooth functions.  To exploit this property we define a
$\Delta R_{jj}$-dependent transformation which turns the actual phase
space pattern into a smoothly dropping curve, let the network learn
this smooth function extremely well, and then undo the transformation
to re-build the local maximum pattern. A simple smoothing function for
our case is
\begin{align}
  f(\Delta R) =
  \begin{cases}
    0 & \text{for } \Delta R < R_- \\[2mm]
    \dfrac{\Delta R - R_-}{R_+ - R_-} & \text{for } \Delta R \in [R_{-},R_+] \\[2mm]
    1 & \text{for }  \Delta R > R_+ \; .
  \end{cases}
\end{align}
The transition region is defined such that it includes the cutoff to ensure non-vanishing weights,
$R_- < R_\text{min} = 0.4$, and its upper boundary is in a stable
phase space regime. In our case we use $R_- = 0.2$ and $R_+ = 1.5$
without much fine-tuning. We also apply this transformation to the
3-jet sample, where all $\Delta R_{jj}$-distribution have similar
challenges, through additional event weights
\begin{align}
  w^\text{(1-jet)} &= 1 \notag \\
  w^\text{(2-jet)} &= f(\Delta R_{j_1,j_2}) \notag \\
  w^\text{(3-jet)} &= f(\Delta R_{j_1,j_2}) f(\Delta R_{j_2,j_3})
                                           f(\Delta R_{j_1,j_3}) \; .
\label{eq:trafo}
\end{align}
After training the INN generator on these modified events we also
enforce the jet separation and set all event weights with $\Delta
R_{jj} < \Delta R_\text{min}$ to zero. The inverse factor compensating
for our magic transformation is then
\begin{align}
  \tilde{f}(\Delta R) =
  \begin{cases}
    0 & \text{for } \Delta R < R_\text{min} \\[2mm]
    \dfrac{R_+ - R_-}{\Delta R - R_-} & \text{for } \Delta R \in [R_\text{min},R_+] \\[2mm]
    1 & \text{for } \Delta R > R_+ 
  \end{cases} \; .
\end{align}
To train the INN generator on weighted data the loss function of
Eq.\eqref{eq:inn_loss} has to be changed to
\begin{align}
  \loss_G = \sum_{i=1}^B
  \left( \frac{\psi(x_i; c_i)}{2} - J(x_i) \right) \; 
  \dfrac{w(x_i)}{\sum_{i=1}^B w(x_i)} \; ,
  \label{eq:inn_loss2}
\end{align}
per batch with size $B$.  Here, the weights are defined in
Eq.\eqref{eq:trafo}, $x_i$ are the latent space vectors, and $J_i$ are
the corresponding logarithms of the Jacobian.  In the right panel of
Fig.~\ref{fig:inn_2d} we see that our network architecture indeed
captures the intricate structure in the jet-jet correlations. The distribution of the resulting event
weights is shown in Fig.~\ref{fig:trafo}. By construction all finite
event weights are above one, and hardly any of them reach values for
more than seven, which implies that these weights can be easily
removed by standard reweighting techniques.

Our magic transformation is similar to a refinement, defined as
per-event modifications of phase space
distributions~\cite{Erdmann:2018kuh}, whereas reweighting uses weights
for individual phase space points or events to improve the agreement
between generator output and
truth~\cite{Diefenbacher:2020rna}. However, our transformation is, by
standard phase-space mapping arguments, counter-intuitive\footnote{As
  a matter of fact, our magic transformation of the density is the
  exact opposite of the standard phase space mapping for Monte Carlo
  integration.}. Instead of removing a leading dependence from a curve
and learning a small but non-trivial difference, we smooth out a
subtle pattern and rely on an outstanding network interpolation to
learn the smoothed-out function better than the original pattern. This
is motivated by the way flow networks learn distributions, which is
more similar to a fit than a combination of local
patterns~\cite{Bellagente:2021yyh}.  The technical disadvantage of the
smoothing transformation is that the generated events are now
weighted, its advantage is that it is very versatile. Another
disadvantage is that it needs to be applied based on an observed
deficiency of the network and does not systematically improve the
training of generative INNs, so below we will try to find alternative
solutions to improve the network performance.

%%%%%%%%%%%%%%%%%%%%%%%%%%%%%%%%%%%%%%%%%%%%%%%%%%%%%%%%%%%%%%%%%%%%%%%%
\subsubsection*{INN-generator benchmark}

%----------------------------------------------------------
\begin{figure}[t]
    \includegraphics[page = 3, width=0.495\textwidth]{plots/2_3_reweighting_trick_1jet}
    \includegraphics[page = 8, width=0.495\textwidth]{plots/2_3_reweighting_trick_1jet} \\
    \includegraphics[page = 4, width=0.495\textwidth]{plots/2_3_reweighting_trick_2jet}
    \includegraphics[page =24, width=0.495\textwidth]{plots/2_3_reweighting_trick_2jet} \\
    \includegraphics[page =35, width=0.495\textwidth]{plots/2_3_reweighting_trick_3jet}
    \includegraphics[page = 1, width=0.495\textwidth]{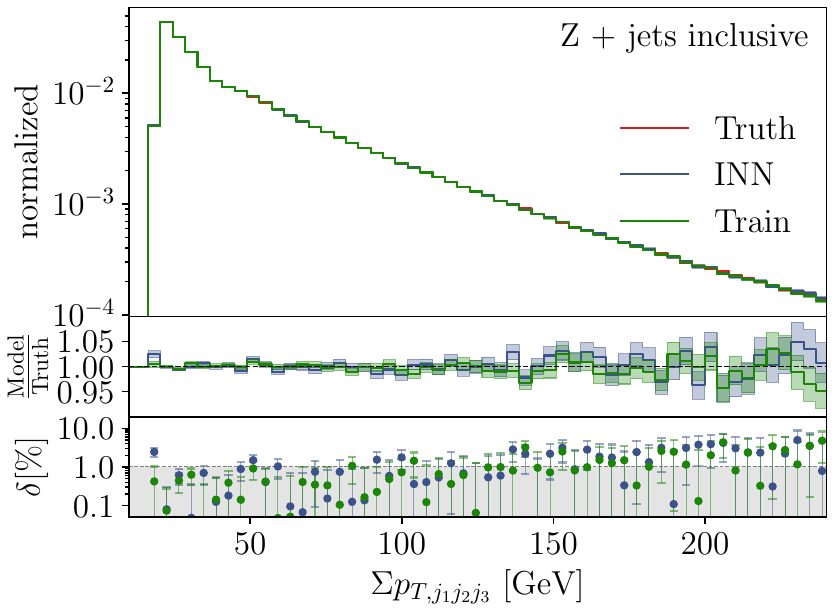}
    \caption{INN distributions for $Z+1$~jet (upper), $Z+2$~jets
      (middle), $Z+3$~jets (lower left) and an inclusive distribution (lower right)
      from a combined $Z+$~jets
      generation. We show weighted events using the magic
      transformation of Eq.\eqref{eq:trafo} to improve the $\Delta R$
      distributions.}
    \label{fig:inn}
\end{figure}
%----------------------------------------------------------

In Fig.~\ref{fig:inn} we show a set of kinematic distributions for our
training data, truth defined as a statistically independent version of
the training sample, and the output of the INN-generator with the
magic transformation of Eq.\eqref{eq:trafo}. We show distributions for
exclusive $Z+ \{ 1,2,3 \}$~jets samples and define the relative
deviation for binned kinematic distributions as
\begin{align}
\delta [\%] = 100 \; \frac{|\text{Model} - \text{Truth}|}{\text{Truth}} \; .
\end{align}
In the top row the final state consists of the $Z$-decay products and
one recoil jet, and we see that the recoil spectrum as well as the
sharp $Z$-mass are learned with high precision. That remains true when
we add a second jet, including the critical $\Delta R_{j_1 j_2}$
correlation discussed above. Finally, adding yet another jet the
network learns the complete set of angular jet-jet
correlations. Looking a the precision of the training sample, which
consists of half of our full data set, we see that at least in the
bulk of the kinematic distribution, the training data set agrees with
truth at the percent level or better. This changes in the kinematic
tails, where the statistical precision of the training data drops
continuously. The level of agreement between the INN-generated events
and truth also reaches the percent level in densely populated phase
space regions, but it is slightly worse than the precision of the
training sample. Also the $Z$-peak even in the 1-jet sample is not
perfectly learned by the INN, which leaves us a little bit of work to
do on the precision side. To confirm that the differemt jet-exclusive
samples are combined correctly, we show the hadronic $H_T$ or scalar
sum of all transverse jet momenta in the lower-right panel. Its
precision nicely tracks that of the different $p_{T,j}$ distributions.

%%%%%%%%%%%%%%%%%%%%%%%%%%%%%%%%%%%%%%%%%%%%%%%%%%%%%%%%%%%%%%%%%%%%%%%%
\section{DiscFlow generator}
\label{sec:gan}

One way to systemically improve and control a precision INN-generator
is to combine it with a discriminator. It is inspired by incredibly
successful GAN applications also in LHC
simulations~\cite{Goodfellow:2014:GAN:2969033.2969125,Creswell2018,Butter:2020qhk}. In
our preliminary studies we never reached a sufficient precision with
established GAN architectures~\cite{Butter:2019cae}, while
INN-generators proved very promising~\cite{Bellagente:2021yyh}.
Compared to reweighting and refinement methods, a GAN-like setup has
the advantage that the generator and discriminator networks already
communicate during the joint training. We will show how such a
discriminator network can be used to improve precision event
generation and then show how a discriminator can be coupled to our INN
generator in a new DiscFlow architecture.

%%%%%%%%%%%%%%%%%%%%%%%%%%%%%%%%%%%%%%%%%%%%%%%%%%%%%%%%%%%%%%%%%%%%%%%%
\subsection{Discriminator reweighting}
\label{sec:gan_disc}

%----------------------------------------------------------
\begin{table}[b!]
  \centering
  \begin{small}
      \begin{tabular}{lcc}
        \toprule
        hyper-parameter 				& value	\\
        \midrule
        LR scheduling                   & Reduce-on-plateau\\
        Starter LR			& $1 \cdot 10^{-2}$	\\
        Epochs                          & 200 \\
        Batch size			& 2048	\\
        Adam $\beta_1$, $\beta_2$       & $0.5$, $0.9$ \\
        Layer type                      & Dense\\
        Number layers                   & 8\\
        Internal size                   & 256\\
        \bottomrule
      \end{tabular}
  \end{small}
  \caption{Training setup and hyperparameters for the discriminator.}
  \label{tab:disc}
\end{table}
%----------------------------------------------------------

%----------------------------------------------------------
\begin{figure}[t]
  \includegraphics[page =  3, width=0.495\textwidth]{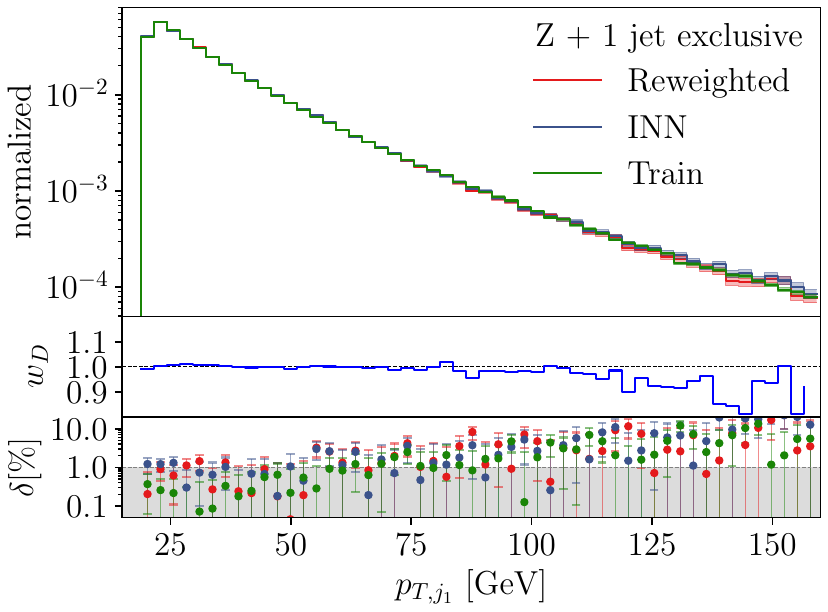}
  \includegraphics[page =  8, width=0.495\textwidth]{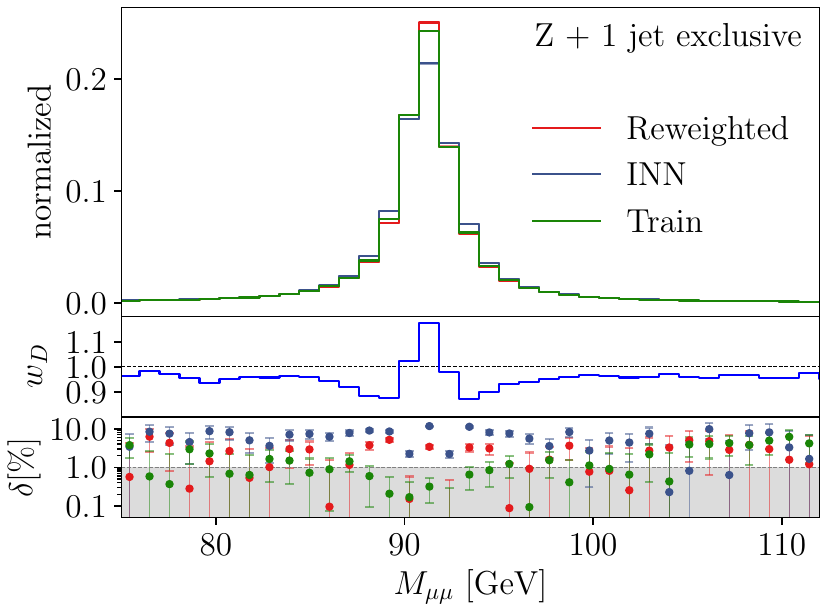}
  \includegraphics[page =  4, width=0.495\textwidth]{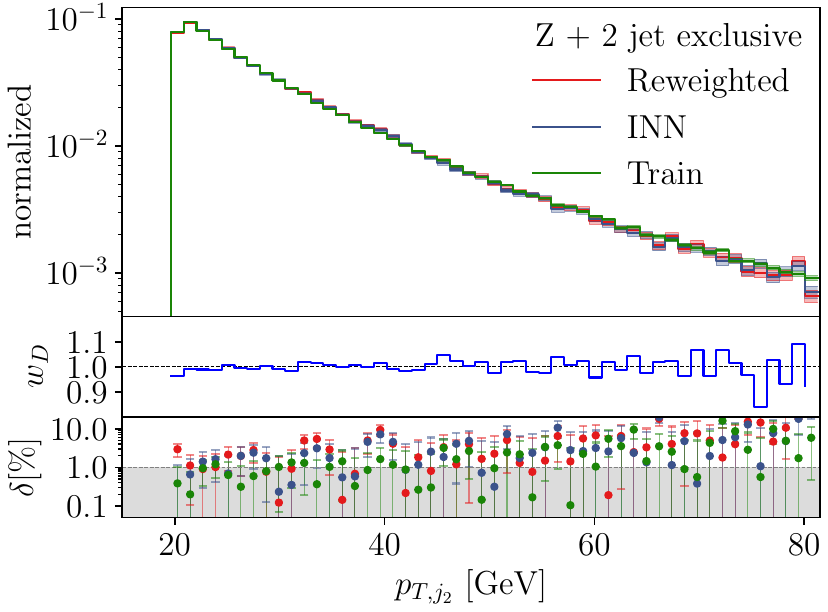}
  \includegraphics[page = 24, width=0.495\textwidth]{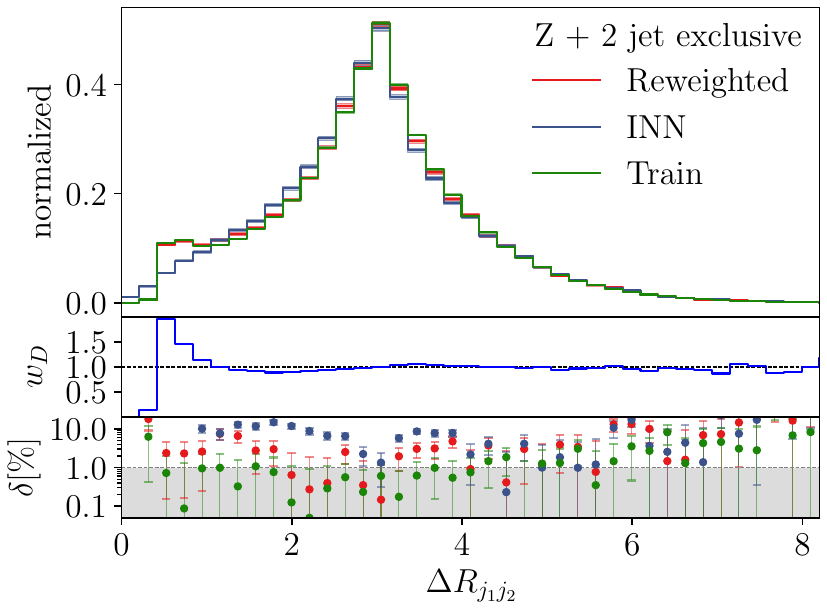}
  \includegraphics[page = 35, width=0.495\textwidth]{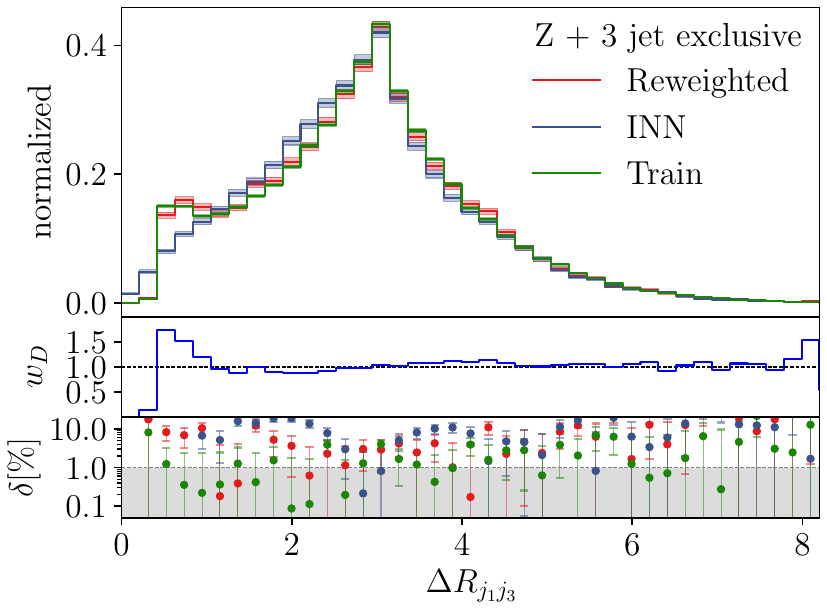}
  \includegraphics[page = 34, width=0.495\textwidth]{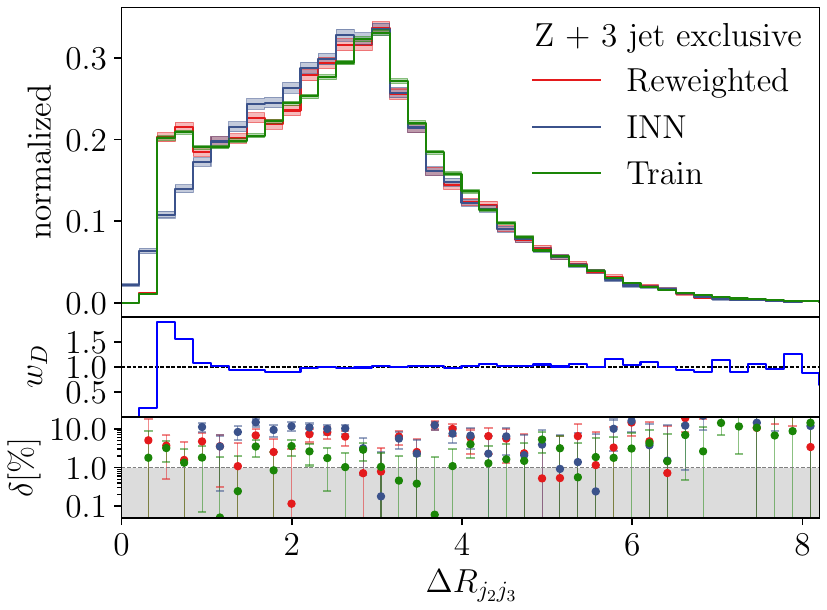}
  \caption{Discriminator-reweighted INN distributions for $Z+1$~jet
    (upper), $Z+2$~jets (middle), and $Z+3$~jets (lower) from a
    combined $Z+$~jets generation. The bottom panels show the
    average correction factor obtained from the discriminator output,
    the INN results without
    reweighting are the same as in Fig.~\ref{fig:inn}, except for
    slightly longer training.}
  \label{fig:reweight}
\end{figure}
%----------------------------------------------------------

Before we train our INN-generator jointly with a discriminator, we
illustrate the power of such a discriminator by training it
independently and reweighting events with the discriminator
output~\cite{Diefenbacher:2020rna}. This requires that our
discriminator output can eventually be transformed into a
probabilistic correction. We train a simple network described in
Tab.~\ref{tab:disc} by minimizing the cross entropy to extract a
probability $D(x_i) \to 0 (1)$ for an identified generator (truth)
event $x_i$.  For a perfect generated sample the discriminator cannot
tell generated events from true events, and the output becomes $D(x_i)
= 0.5$ everywhere.  Using this discriminator output we define the
event weight
\begin{align}
  w_D(x_i) = \frac{D(x_i)}{1-D(x_i)} \to \frac{P_\text{data}(x_i)}{P(x_i)} \; .
\label{eq:reweight}
\end{align}
In the conventions of Eq.\eqref{eq:inn_loss_cont} $w_D$ approximates 
the ratio of true over generated phase space densities, so we can use
it to reweight each event such that it reproduces the true kinematic
distributions at the level they are encoded in the discriminator.

To see how precisely this kind of discriminator works we use the
standard INN generator from Sec.~\ref{sec:prec_arch}. We omit the
magic transformation described in Eq.\eqref{eq:trafo}, to define a
challenge for the discriminator. For each jet-multiplicity of the cINN
model, we train a discriminative model in parallel to the generative
model, but for now without the two networks communicating with each
other. The input to the three distinct discriminator networks, one per
multiplicity, are the usual observables $p_T, \eta, \phi$, and $m$ of
Eq.\eqref{eq:def_obs} for each final-state particle. We explicitly include a set
of correlations known to challenge our naive INN generator
and train the discriminator
\begin{align}
  \loss_D = -\sum_i^B\log(1-D(x_{i, \text{gen}}))-\sum_i^B\log(D(x_{i, \text{data}})
\end{align}
with generated vectors extended depending on the jet multiplicity
\begin{align}
  x_i = \{p_{T,j},\eta_j,\phi_j, M_j \}
  \cup \{M_{\mu \mu}\}
  \cup\{\Delta R_{2,3}\}
  \cup \{\Delta R_{2,4}, \Delta R_{3,4}\} \; .
\end{align}
and corresponding training vectors $x_{i, \text{data}}$.

In Fig.~\ref{fig:reweight} we show sample kinematic distributions for
the $Z+ \{1,2,3\}$ jet final states. Truth is defined as the
high-statistics limit of the training data.  The INN events are
generated with the default generator, without the magic transformation
of Eq.\eqref{eq:trafo}, so they are unweighted events. The reweighted
events are post-processed INN events with the average weight per bin shown in
the second panel. While for some of the shown distribution a flat
dependence $w_D = 1$ indicates that the generator has learned to
reproduce the training data to the best knowledge of the
discriminator, our more challenging distributions are significantly
improved by the discriminator. That includes the reconstructed
$Z$-mass as well as the different $\Delta R_{jj}$-distributions.

Comparing the discriminator-reweighted performance to the magic
transformation results in Fig.~\ref{fig:inn}, reproduced as the blue
lines in Fig.~\ref{fig:reweight}, we see that the tricky distributions
like $\Delta R_{j_1 j_2}$ or $\Delta R_{j_1 j_3}$ are further improved
through the reweighting over their entire range.  For the comparably
flat $p_T$-distributions the precision of the reweighted events is
becoming comparable to the training statistics, both for the bulk of
the distribution and for the sparsely populated tails.  Of all
kinematic distributions we checked, the vector sum of all hard
transverse momenta of the 5-object final state is the only
distribution where the naive INN-generator only learns the phase space
distribution only at the 10\% level. Also those are corrected fine by
the discriminator reweighting.

While the discriminator reweighting provides us with an architecture
that learns complex LHC events at the percent level or at the level
of the training statistics, it comes with the disadvantage of
generating weighted events and does not use the opportunity for the
generator and discriminator to improve each other. Both of these open
questions will be discussed in the next architecture.

%%%%%%%%%%%%%%%%%%%%%%%%%%%%%%%%%%%%%%%%%%%%%%%%%%%%%%%%%%%%%%%%%%%%%%%%
\subsection{Joint training}
\label{sec:gan_joint}

%----------------------------------------------------------
\begin{figure}[t]
  \centering
  \includegraphics[width=0.5\textwidth]{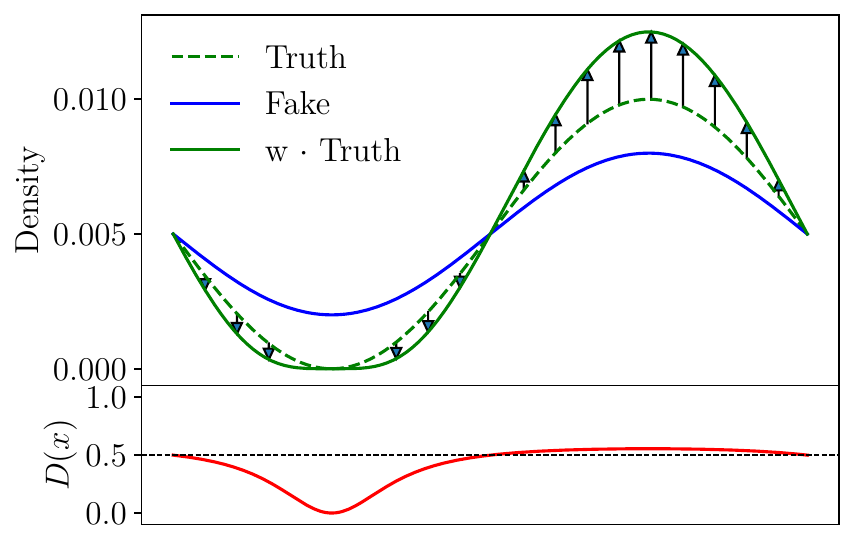}
  \caption{Illustration of the DiscFlow method. Weights computed by
    the discriminator shift the reference (true) density downwards
    whenever the generator (fake) distribution overshoots and
    vice-versa. This way the deviations of the to-be-trained generator
    density are over-exaggerated.}
  \label{fig:WeightedTrueDist}
\end{figure}
%----------------------------------------------------------

After observing the benefits from an additional discriminator network,
the question is how we can make use of this second network most
efficiently. If it is possible to train the discriminator and
generator network in parallel and give them access to each other, a
joint GAN-like setup could be very
efficient~\cite{grover2018flowgan}. Unfortunately, we have not been
able to reach the required Nash equilibrium in an adversarial training
for our specific INN setup. Instead, one of the two players was always
able to overpower the other.

%----------------------------------------------------------
\begin{figure}[t]
  \includegraphics[page =  3, width=0.495\textwidth]{plots/3_2_INN_joint3}
  \includegraphics[page =  8, width=0.495\textwidth]{plots/3_2_INN_joint3}
  \includegraphics[page =  1, width=0.495\textwidth]{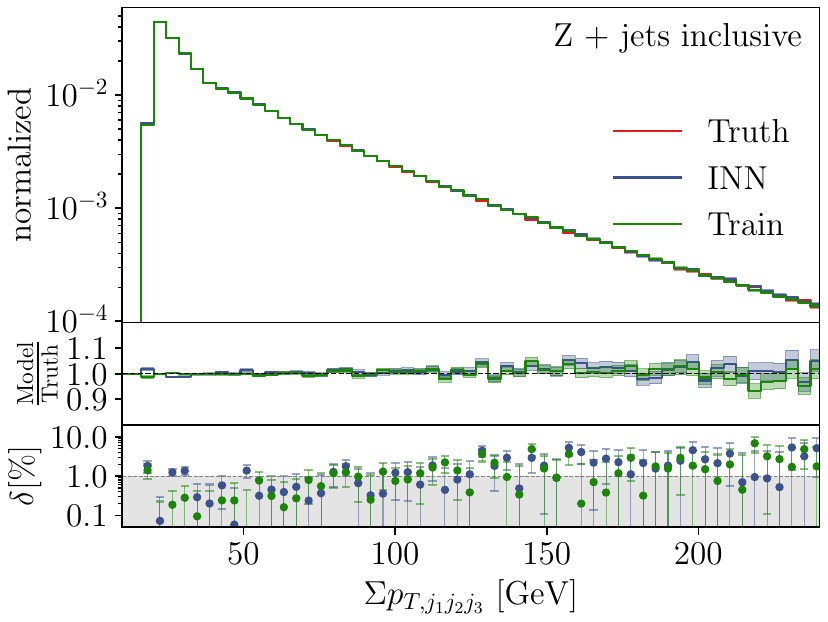}
  \includegraphics[page = 35, width=0.495\textwidth]{plots/3_2_INN_joint5}
  \caption{DiscFlow distributions for $Z+1$~jet, $Z+3$~jets,
    and an inclusive distribution
    from a combined $Z+$~jets generation after joint
    generator--discriminator training.}
    \label{fig:discflow}
\end{figure}
%----------------------------------------------------------

Instead of relying on a Nash equilibrium between the two
competing network architectures we can avoid a two-part loss functions
entirely and incorporate the discriminator information into the
generator loss of Eq.\eqref{eq:inn_loss} through the event weight
function $w_D(x)$ of Eq.\eqref{eq:reweight},
\begin{align}
  \loss_\text{DiscFlow}
  &=
  - \sum_{i=1}^B \; w_D(x_i)^\alpha \; \log \frac{P(x_i;c_i)}{P_\text{data}(x_i;c_i)} \notag \\
%  &\approx
%  \int dx \; \underbrace{w_D(x)^\alpha P_\text{data}(x)}_\text{reweighted truth} \;
%  \; \log \frac{P(x;c)}{P_\text{data}(x;c)} \notag \\
  &\approx
  - \int dx \; \frac{P_\text{data}^{\alpha+1}(x)}{P^\alpha(x)} \;
  \log \frac{P(x)}{P_\text{data}(x)} \notag \\
  &=
  - \int dx \; \left( \frac{P_\text{data}(x)}{P(x)} \right)^{\alpha+1} 
  \; P(x) \log P(x)
  + \int dx \; \left( \frac{P_\text{data}(x)}{P(x)} \right)^\alpha 
  \; P_\text{data}(x) \log P_\text{data}(x) \notag \\
  &=
    - \XLangle \left( \frac{P_\text{data}(x)}{P(x)} \right)^{\alpha+1} \log P(x) \XRangle_{P}
  + \XLangle \left( \frac{P_\text{data}(x)}{P(x)} \right)^\alpha \log P_\text{data}(x) \XRangle_{P_\text{data}}
  \; , 
  \label{eq:inn_loss3}
\end{align}
with an appropriately defined expectation value. For the continuum
limit we omit the conditional argument and assume a perfectly trained
discriminator. Note that in our simple DiscFlow setup the
discriminator weights $\omega_D \approx P_\text{data}(x)/P(x)$ do not
have gradients with respect to the generative model parameters, so
only the first term in the last line contributes to the
optimization. This term corresponds to the negative log-likelihood of
training samples drawn from the weighted truth distribution.
%
%\begin{align}
%  \loss_\text{DiscFlow}
%  &=
%%  &= w_D(x)^\alpha \mathcal{L}_\text{Latent}(x)
%  \sum_{i=1}^B w_D(x_i)^\alpha \left( \frac{\psi(x_i; c_i)^2}{2} - \log J(x_i) \right) \notag \\
%  &\approx
%  \int dx \; \underbrace{w_D(x)^\alpha P(x)}_\text{reweighted truth} \;
%  \left( \frac{\psi(x; c)^2}{2} - \log J(x) \right) 
%  \; .
%  \label{eq:inn_loss3}
%\end{align}
%
The hyperparameter $\alpha$ determines the impact of the discriminator
output, and we introduce an additional discriminator dependence as
\begin{align}
  \alpha = \alpha_0 \; \left| \frac{1}{2} - D(x) \right| \; .
\end{align}
During training we increase $\alpha_0$ linearly to enhance the impact
of the reweighting factor, while the improved training will drive the
discriminator to $D(x) \to 1/2$. This functional form for $\alpha$ is
the simplest way of combining the two effects towards a stable result.

From Eq.\eqref{eq:inn_loss3} we see that our modified loss is equivalent
to training on a shifted reference distribution. In
Fig.~\ref{fig:WeightedTrueDist} we illustrate what happens if the
generator populates a phase space region too densely and we reduce the
weight of the training events there. Conversely, if a region is too
sparsely populated by the generator, increased loss weights amplify
the effect of the training events.  Our new discriminator--generator
coupling through weights has the advantage that it does not require a
Nash equilibrium between two competing networks, so the discriminator
can no longer overpower the generator.  As the generator converges
towards the true distribution, the discriminator will stabilize as
$w_D(x) \rightarrow 1$, and the generator loss will approach its
unweighted global minimum.

%----------------------------------------------------------
\begin{figure}[t]
  \includegraphics[page =  3, width=0.495\textwidth]{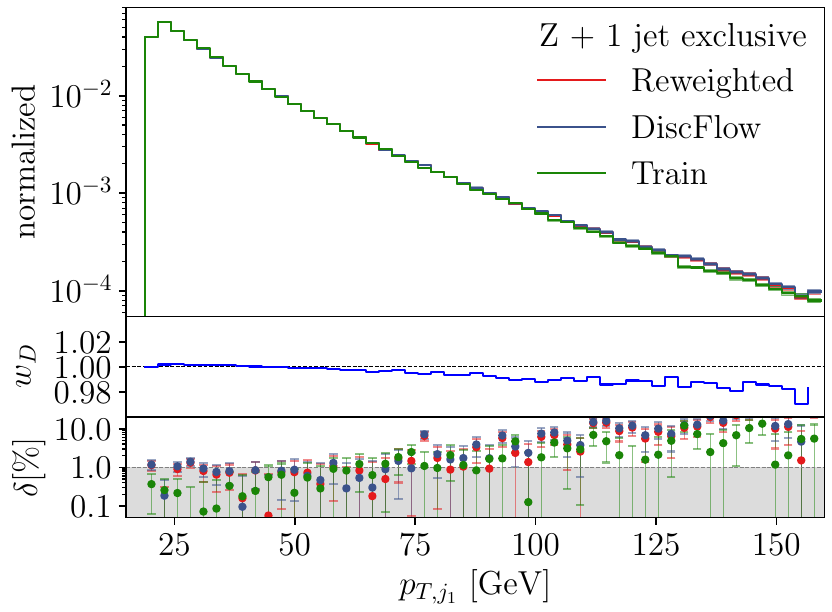}
  \includegraphics[page =  8, width=0.495\textwidth]{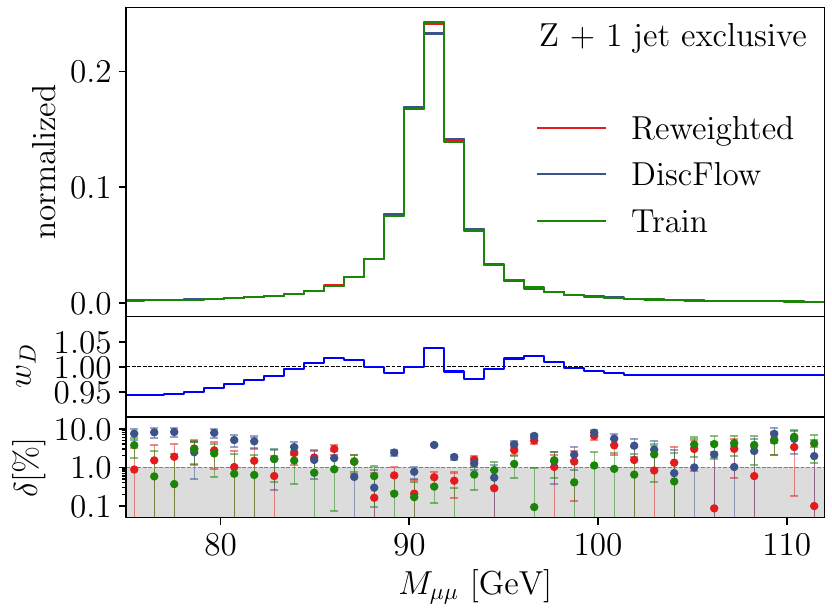}
  \includegraphics[page =  4, width=0.495\textwidth]{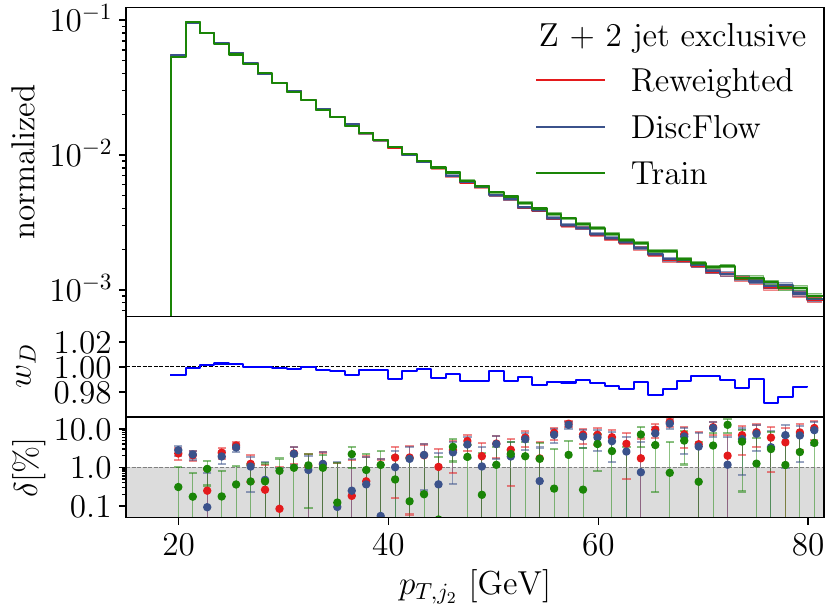}
  \includegraphics[page = 24, width=0.495\textwidth]{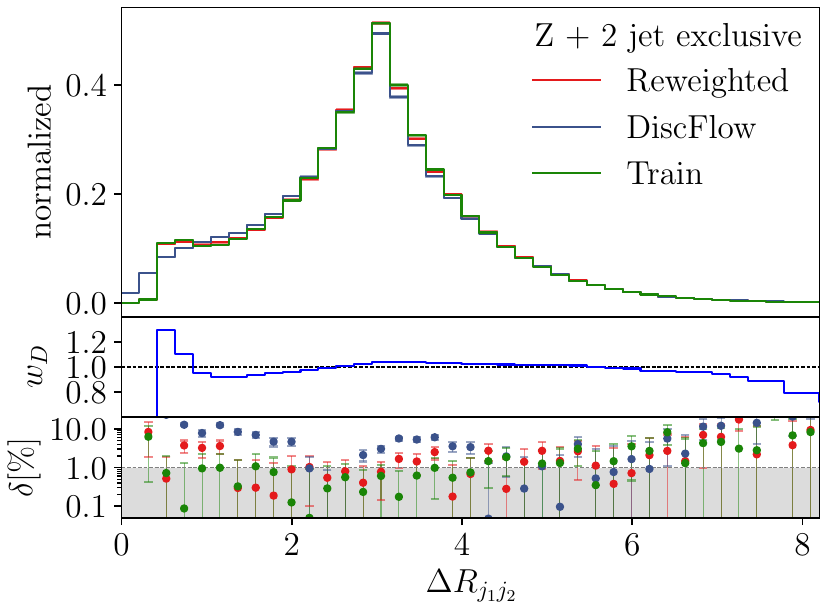}
  \includegraphics[page = 35, width=0.495\textwidth]{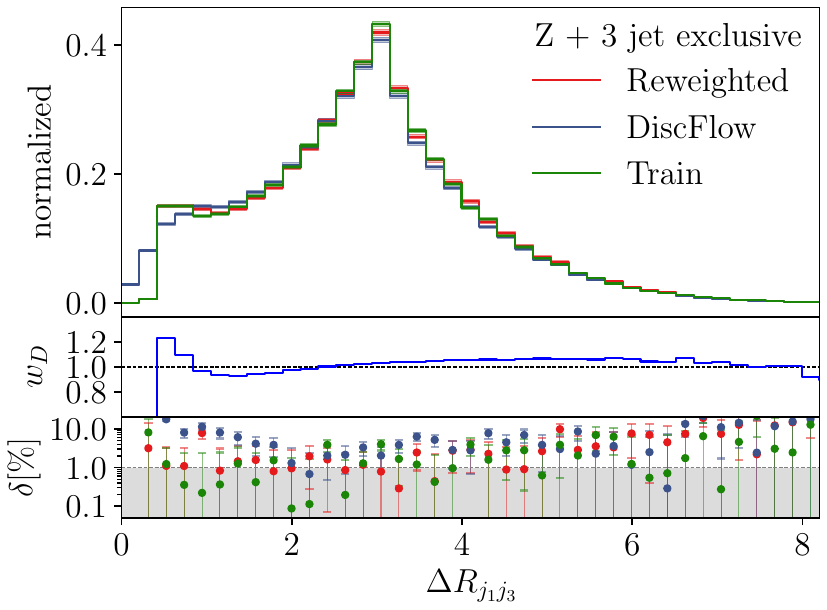}
  \includegraphics[page = 34, width=0.495\textwidth]{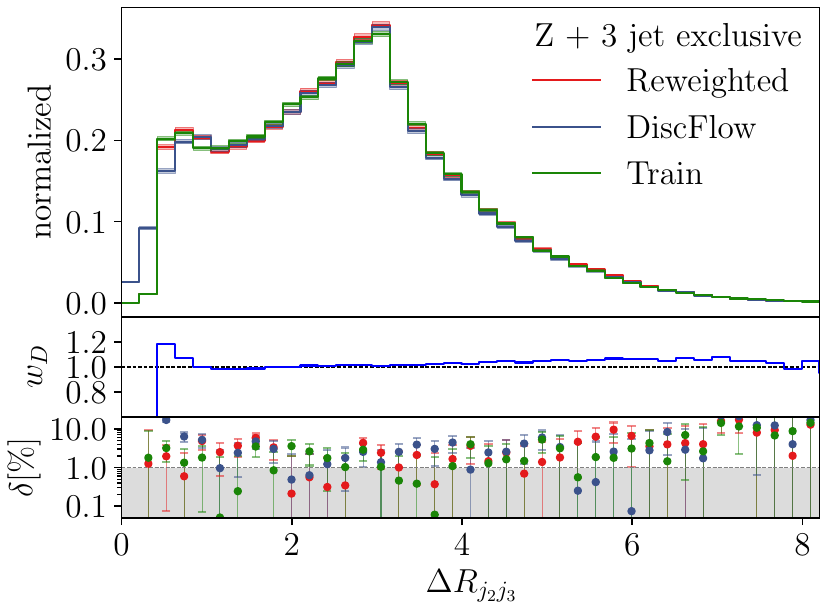}
  \caption{Discriminator-reweighted DiscFlow distributions for
    $Z+1$~jet (upper), $Z+2$~jets (middle), and $Z+3$~jets (lower)
    from a combined $Z+$~jets generation. The bottom panels show the
    average correction factor obtained from the discriminator output.
    The DiscFlow results for joint generator--discriminator training
    without reweighting are the same as in Fig.~\ref{fig:discflow}.}
  \label{fig:discflow_re}
\end{figure}
%----------------------------------------------------------

When training the two DiscFlow networks jointly, we split the batches
per epoch equally between both networks, training each network on a
separate subset of the training data.  To increase the stability, we
start by training the generator and the separate discriminators for
the different jet multiplicities separately and only combine them to a
stable joint training once all networks are pre-trained.

In Fig.~\ref{fig:discflow} we show the performance of the DiscFlow
setup to our $Z$+jets benchmark process. First, we see that in the
bulk of the flat distributions like $p_{T,j}$ the generator reproduces
the correct phase space density almost at the level of the training
statistics. Comparing the results to Fig.~\ref{fig:inn} and
Fig.~\ref{fig:reweight} we see a comparable, possibly improved,
performance of the joint training. The non-negligible density of
generated events below the cut at $\Delta R = 0.4$ shows that the
DiscFlow method is only effective in phase space regions populated by
training data.  These results indicate that the joint training of the
generator with a discriminator corrects the invariant mass and all
other tricky distribitions almost to the level of the training
statistics, but with unweighted events, unlike for the magic
transformation in Fig.~\ref{fig:inn} and the explicit reweighting in
Fig.~\ref{fig:reweight}.

In the ideal AI-world we assume that after successful joint training
the discriminator will have transferred all of its information into
the generator, such that $D(x)=0.5$ at any point of phase space. In
reality, this is not at all guaranteed. We know from
Fig.~\ref{fig:reweight} that the discriminator can learn the $\Delta
R$ features very well, so we combine the joint training and
discriminator reweighting strategies to ensure that we extract the
full performance of both networks. In Fig.~\ref{fig:discflow_re} we
show the same training results as in Fig.~\ref{fig:discflow}, but
reweighted with $w_D$.  We see that the reweighting leads to a small
correction of the $M_{\mu \mu}$-distribution and a sizeable correction
to the $\Delta R_{jj}$ features close to the jet separation cut.
Because of the way we provide the event input, we note that the
transverse momentum conservation would become the next challenge after
mastering $M_{\mu \mu}$ and $\Delta R_{jj}$. For all other observables
our reweighted DiscFlow network indeed reproduces the true kinematic
distributions at the percent level provided by the training
statistics.

%----------------------------------------------------------
\begin{figure}[t]
  \centering
  \includegraphics[width=0.495\textwidth, page=3]{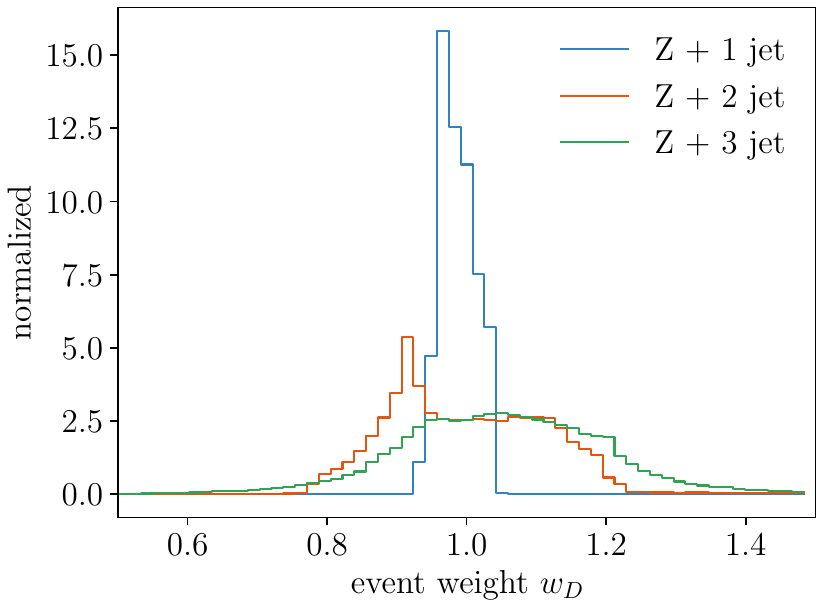}
  \caption{Distribution over the weights $w_D$ computed over the
    entire, not marginalized phase space.}
  \label{fig:weights}
\end{figure}
%----------------------------------------------------------

While in Fig.~\ref{fig:discflow_re} we see that the correction factor
obtained from the discriminator shows the agreement of training events
and simulated events, it
is crucial that we search the fully exclusive phase space for
systematic deviations between training and simulated events. In
Fig.~\ref{fig:weights} we histogram all event weights $w_D(x_i)$ for
$Z+$~jets production.  For the high-statistics $Z+1$~jet sample the
correction weights are at most at the percent level. The fact that our
generator only learns the phase space density and not the total rates
allows for a slight bias in the event weight distributions.  For the
bulk of the kinematic distributions the bin-wise correction in
Fig.~\ref{fig:discflow_re} is still slightly smaller than the weights
shown here, which means that some of the corrections are simply noise.
The width of the weight distribution increases for higher jet
multiplicities, simply reflecting the drop in training statistics.
Combining Fig.~\ref{fig:weights} and Fig.~\ref{fig:discflow_re} allows
us to trace the large weights $w_D$ to critical phase space regions,
like the lower tail of the $M_{\mu \mu}$-distribution for $Z+1$~jet or
$\Delta R_{j j} \lesssim 0.5$ for $Z+2/3$~jets.

%%%%%%%%%%%%%%%%%%%%%%%%%%%%%%%%%%%%%%%%%%%%%%%%%%%%%%%%%%%%%%%%%%%%%%%%
\section{Uncertainties and control}
\label{sec:uncert}

After introducing our precision generator architecture in
Sec.~\ref{sec:prec} and extending it to a discriminator--generator
architecture for control in Sec.~\ref{sec:gan}, the last item on our
list of LHC tasks is a comprehensive treatment of uncertainties. A
proper uncertainty treatment has been discussed for instance for
regression or classification
networks~\cite{Bollweg:2019skg,Kasieczka:2020vlh,Nachman:2019dol},
while for generative networks there exists only a first study on how
to use and interpret Bayesian INNs~\cite{Bellagente:2021yyh}. In this
final section we discuss how different uncertainties on generated
events can be extracted using a Bayesian generator network, a
conditional sampling using simulated uncertainties, and the
discriminator introduced in the previous section. Each of these
handles allows us to control certain kinds of uncertainties, and in
combination they allow us to extract a meaningful uncertainty map over
phase space.

%%%%%%%%%%%%%%%%%%%%%%%%%%%%%%%%%%%%%%%%%%%%%%%%%%%%%%%%%%%%%%%%%%%%%%%%
\subsection{Bayesian network}
\label{sec:uncert_bayes}

The simple idea behind Bayesian networks is to replace trained network
weights by trained distributions of network weights. If we evaluate
the network by sampling over these distributions, the network output
will be a central value of the numerically defined function and an
uncertainty
distribution~\cite{bnn_early,bnn_early2,bnn_early3}. Because general
MCMC-methods become expensive for larger networks, we rely on
variational inference to generate the weight
distributions~\cite{blei2017variational}. More specifically, we rely
on a Gaussian approximation for the network weight distribution and
learn the mean and the standard deviation instead of just one value in
a deterministic network. Because of the non-linear nature of the
network the output does not have a Gaussian uncertainty
distribution~\cite{Kasieczka:2020vlh}. Our Bayesian INN (BINN) follows
the same setup as our deterministic INN-generator in
Sec.~\ref{sec:prec_arch}, converted to the Bayesian setup following
Ref.~\cite{Bellagente:2021yyh}.

For a Bayesian generative network we supplement the phase space
density $p(x )$, encoded in the density of unweighted events, with an
uncertainty map $\sigma (x )$ over the same phase space. To extract
the density we bin events in a histogram for a given observable and
with finite statistics. Focussing on one histogram and omitting the
corresponding phase space argument $x$ the expected number of events
per bin is
\begin{align}
  \mu \equiv \langle n \rangle = \sum_n n P_N(n) \; ,
\label{eq:nexpected1}
\end{align}
with $P_N(n)$ given by the binomial or Poisson probability of
observing $n$ events in this bin. This event count should be the mean
of the BINN distribution, defined by sampling from the distribution $q(\theta)$
over the network weights $\theta$,
\begin{align}
  \langle n \rangle
  = \int d\theta \; q(\theta) \sum_n n P_N(n|\theta)
  \equiv \int d\theta \; q(\theta) \; \langle n \rangle_\theta \; .
\label{eq:nexpected2}
\end{align}
Following the same argument as in Ref.~\cite{Kasieczka:2020vlh} we can
compute the standard deviation of this sampled event count and split
it into two terms,
\begin{align}
  \sigma_\text{tot}^2
  &= \langle (n - \langle n \rangle)^2 \rangle \notag \\
%  &= \int d\theta \: q(\theta) \langle (n - n\langle n \rangle)^2 \rangle_\theta \\
%  &= \int d\theta \: q(\theta) \langle n^2 - 2 n \langle n \rangle +
%  \langle n \rangle^2 \rangle_\theta \\
  &= \int d\theta \: q(\theta) \left[ \langle n^2 \rangle_\theta -
    2 \langle n \rangle_\theta \langle n \rangle + \langle n \rangle^2 \right] \notag \\
  &= \int d \theta \: q(\theta) \left[ \langle n^2 \rangle_\theta -
    \langle n \rangle_\theta^2 + (\langle n \rangle_\theta - \langle n \rangle)^2 \right] 
  \equiv \sigma_\text{stoch}^2 + \sigma_\text{pred}^2 \; .
\label{eq:sigma_tot}
\end{align}
The first contribution to the uncertainty is the variance of the Poisson distribution,
\begin{align}
  \sigma_\text{stoch}^2
  = \int d\theta \: q(\theta)
  \left[ \langle n^2 \rangle_\theta - \langle n \rangle_\theta^2 \right] 
%  = \int d\theta \: q(\theta) \langle n \rangle_\theta
  = \langle n \rangle \; .
\label{eq:sigma_stoch}
\end{align}
Even if the network is perfectly trained and $q(\theta)$ turns into a
delta distribution, it does not vanish, because it describes the
stochastic nature of our binned data set. The second term,
\begin{align}
  \sigma_\text{pred}^2 = \int d\theta \: q(\theta)
  \left[ \langle n \rangle_\theta - \langle n \rangle \right]^2 \; ,
\label{eq:sigma_pred}
\end{align}
captures the deviation of our network from a perfectly trained
network, where the widths of the network weights vanish.

Moving from a binned to a continuous distribution we can transform our
results into the density and uncertainty maps over phase space,
as introduced in Ref.~\cite{Bellagente:2021yyh}. Assuming $\langle n
\rangle \propto p(x)$, with an appropriate proportionality factor and
a continuous phase space variable $x$, Eqs.\eqref{eq:nexpected2}
and~\eqref{eq:sigma_pred} turn into
\begin{align}
p(x) &= \int d\theta \; q(\theta) \; p(x |\theta) \notag \\ 
\sigma_\text{pred}^2(x) &= \int d\theta \; q(\theta) \; \left[ p(x |\theta) - p(x) \right]^2 \; .
\end{align}

\begin{figure}[t]
  \includegraphics[width=0.495\textwidth]{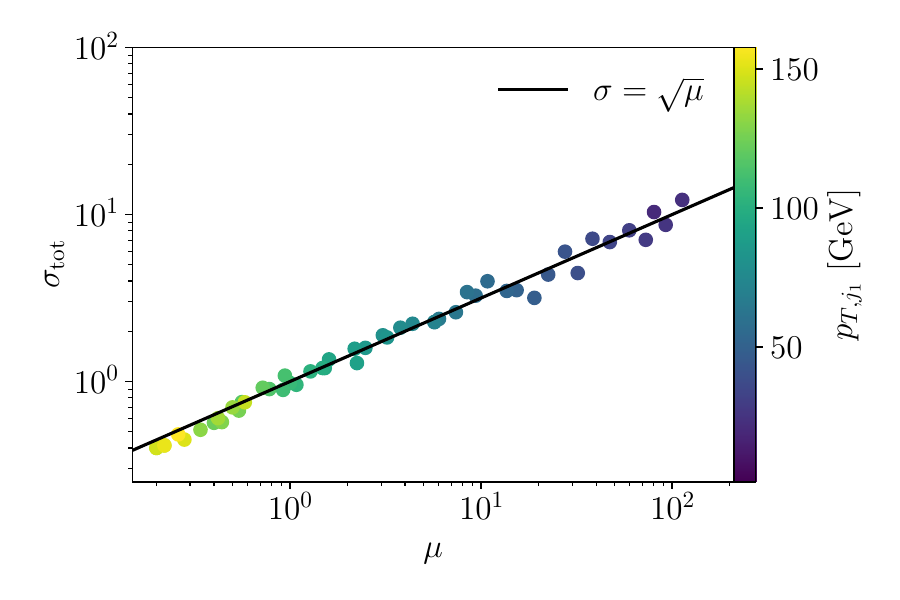}
  \includegraphics[width=0.495\textwidth]{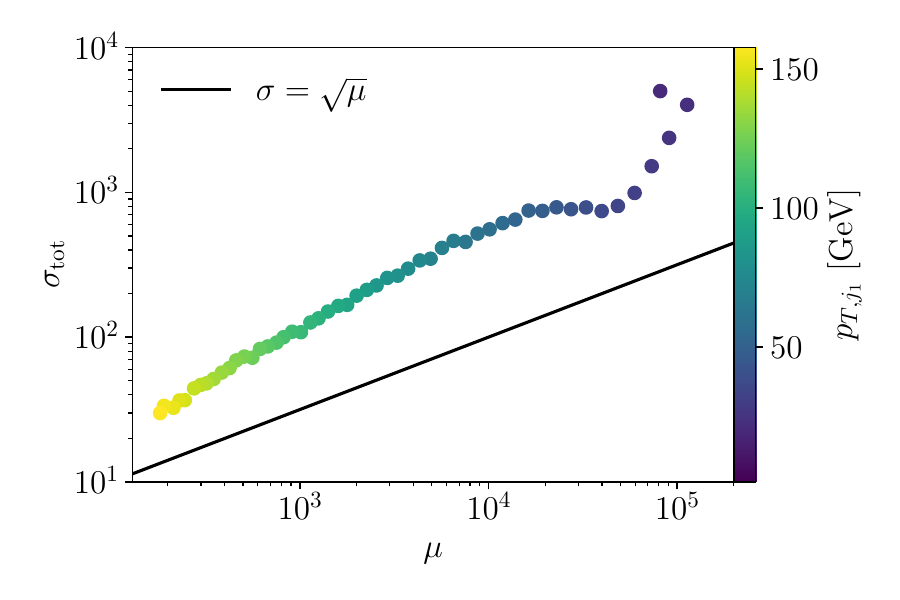}
  \caption{Correlation between event count and BINN uncertainty for
    1000 (left) and 1M (right) generated events. The diagonal like
    defines the Gaussian scaling for a statistically limited sample.}
    \label{fig:binn_sigma_vs_mu}
\end{figure}

To estimate $\sigma_\text{tot}$, we sample $\theta$ and $n$ from
their underlying distributions and compute $\langle n \rangle$. In
practice, we draw weights $\theta$, generate $N$ events with those
weights, histogram them for the observable of interest, extract $n$
per bin. Because the INN-generator is very fast, we can repeat this
process to compute the standard deviation. To see the effect of the
different contributions to the BINN uncertainty we illustrate the
correlation between the event count and $\sigma_\text{tot}$ for
$Z+1$~jet events in Figure \ref{fig:binn_sigma_vs_mu}, with the
$p_{T,j}$-distribution described by 60
bins. Each of these bins corresponds to a dot in the figure. As long
as our sampling is limited by the statistics of the generated events
we find the expected Poisson scaling $\sigma \propto \sqrt{\mu}$,
corresponding to the contribution $\sigma_\text{stoch}$. For larger
statistics, $\sigma_\text{stoch}$ becomes relatively less important,
and the actual predictive uncertainty of the BINN takes over,
$\sigma_\text{tot} \approx \sigma_\text{pred}$.

%----------------------------------------------------------
%----------------------------------------------------------

%----------------------------------------------------------
\begin{figure}[t]
    \includegraphics[width=0.49\textwidth,page= 3]{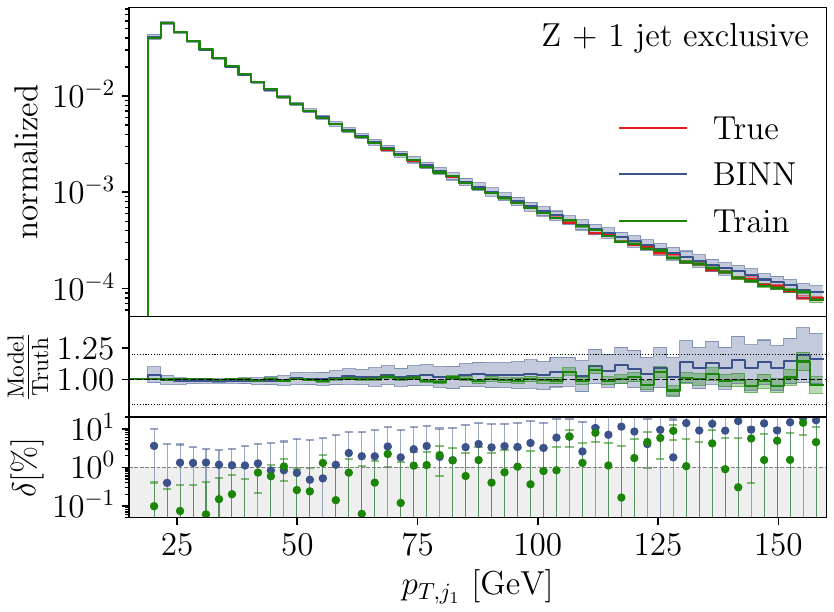}
    \includegraphics[width=0.49\textwidth,page= 8]{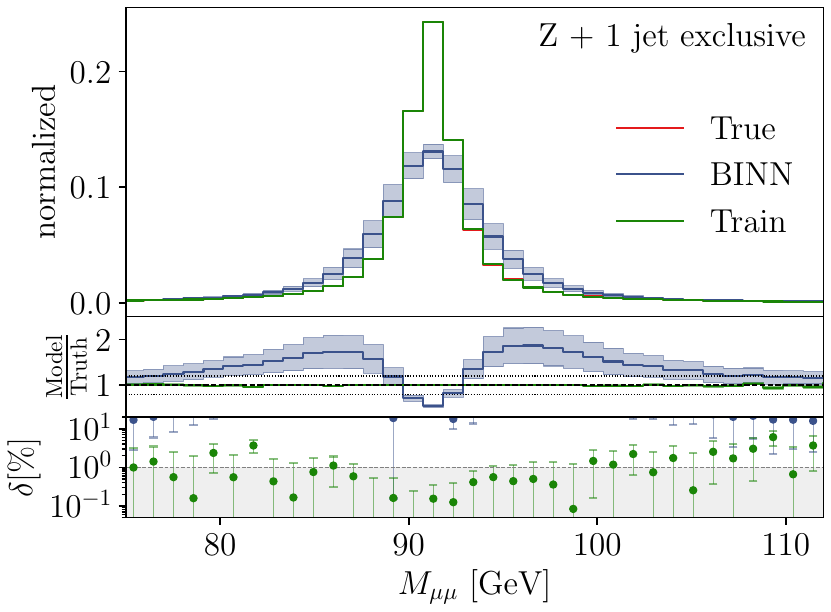}
    \includegraphics[width=0.49\textwidth,page= 4]{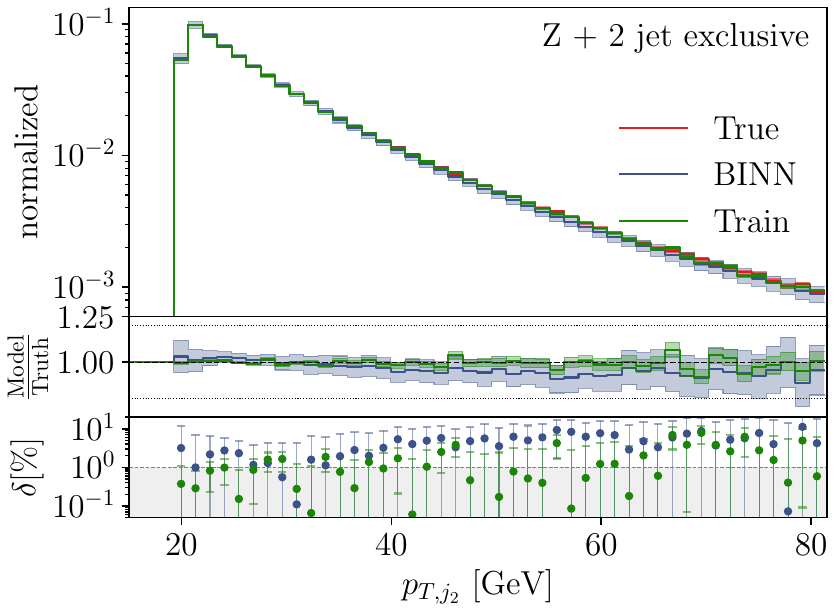}
    \includegraphics[width=0.49\textwidth,page=24]{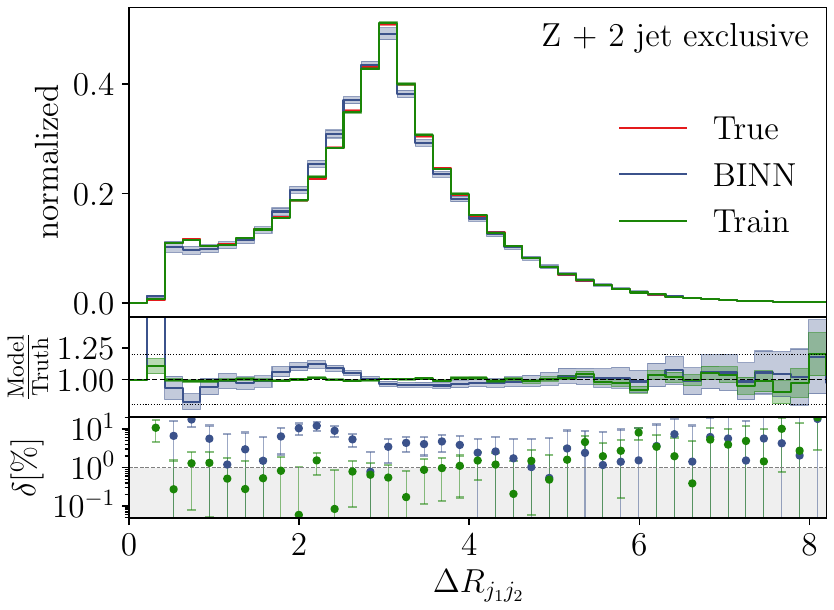}
    \includegraphics[width=0.49\textwidth,page=35]{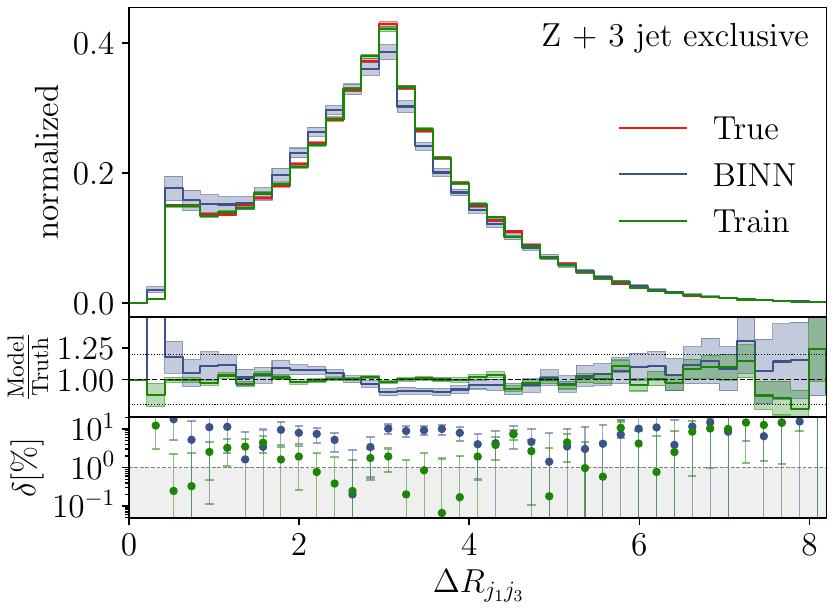}
    \includegraphics[width=0.49\textwidth,page=34]{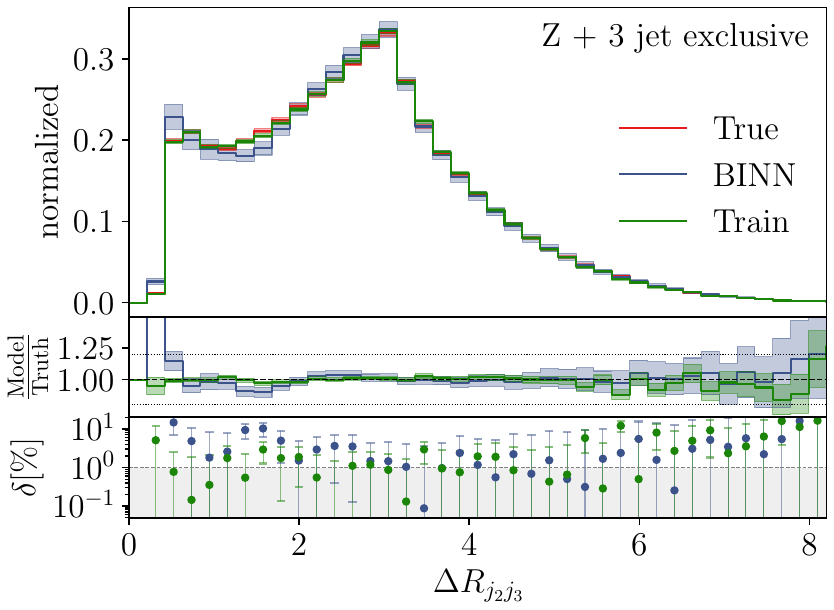}
    \caption{BINN densities and uncertainties for $Z+1$~jet
      (upper), $Z+2$~jets (middle), and $Z+3$~jets (lower) from a
      combined $Z+$~jets generation. The architecture and training
      data correspond to the deterministic network results shown in
      Fig.~\ref{fig:inn}, including the magic transformation of
      Eq.\eqref{eq:trafo}.}
    \label{fig:binn}
\end{figure}
%----------------------------------------------------------

%%%%%%%%%%%%%%%%%%%%%%%%%%%%%%%%%%%%%%%%%%%%%%%%%%%%%%%%%%%%%%%%%%%%%%%%
\subsubsection*{Sources of uncertainties}

By construction, Bayesian networks capture the effects of limited
training statistics and non-perfect training. If we control the truth
information and can augment the training data, a Bayesian network can
also propagate the effects of systematic biases, systematic
uncertainties, or noise into the network
output~\cite{Bollweg:2019skg,Kasieczka:2020vlh}. For generative
networks, the Bayesian network is ideally suited to understand the way
the network learns the phase space density by following the density
map it learns in parallel~\cite{Bellagente:2021yyh}. As a side remark,
we can use this information to track the learning of the BINN for
our $Z$+jets events. We find that the network first learns the
$p_T$-distributions of the different final-state particles quite
precisely, before it targets the angular correlations. This explains
why small features of the $\Delta R$-distributions are the hardest to
learn, because they arise only for the correlation of the $\Delta
\eta$ and $\Delta \phi$ observables. Correspondingly, we find that
one way of improving the performance on the angular correlation is to
apply noise specifically to the $p_T$-distributions. On the other
hand, the magic transformation of Eq.\eqref{eq:trafo} turns out to be
the more efficient solution to this problem, so we also apply it to
the BINN.

When modelling different uncertainties, the problem with augmented
training data for generative networks is that their training is,
strictly speaking, unsupervised. We do not have access to the true
density distribution and have to extract it by binning event
samples. This means that additional noise will only be visible in the
BINN uncertainty if it destabilizes the training altogether. Other
data augmentation will simply lead to a different target density,
overriding the density encoded in the original set of events. This is
why in the following we will discuss training statistics and
stability, and postpone the description of systematics in generative
network training to Sec.~\ref{sec:uncert_cond}.

In Fig.~\ref{fig:binn} we show the uncertainty $\sigma_\text{tot}
\approx \sigma_\text{pred}$ given by the BINN for a Bayesian version
of the network introduced in Sec.~\ref{sec:prec_arch}, including the
magic transformation for the $\Delta R$-distributions. As before, we
see that the network learns the phase space density very precisely for
simple kinematic distributions like $p_{T,j_1}$. The slightly worse
performance compared to the deterministic network in
Fig.~\ref{fig:binn} is due to the increased training effort required
by the larger network. The extracted uncertainties for $p_{T,j_1}$ and
$p_{T,j_2}$ for instance in the bulk reflect the lower statistics of
the $Z+2$~jet training sample compared to $Z+1$~jet.  The narrow
$M_{\mu \mu}$-distribution challenges the uncertainty estimate in that
the network learns neither the density nor the uncertainty very
precisely~\cite{Bellagente:2021yyh}. This limitation will be overcome
once the network learns the feature in the density properly. For the
different $\Delta R$-distributions we see that the network learns the
density well, thanks to the magic transformation of
Eq.\eqref{eq:trafo}. Therefore, the network also reports a comparably
large uncertainty in the critical phase space regions around $\Delta
R_{ij} = 0.4~...~1$.

%%%%%%%%%%%%%%%%%%%%%%%%%%%%%%%%%%%%%%%%%%%%%%%%%%%%%%%%%%%%%%%%%%%%%%%%
\subsubsection*{Effect of training statistics}

%----------------------------------------------------------
\begin{figure}[t]
    \includegraphics[width=0.495\textwidth,page= 2]{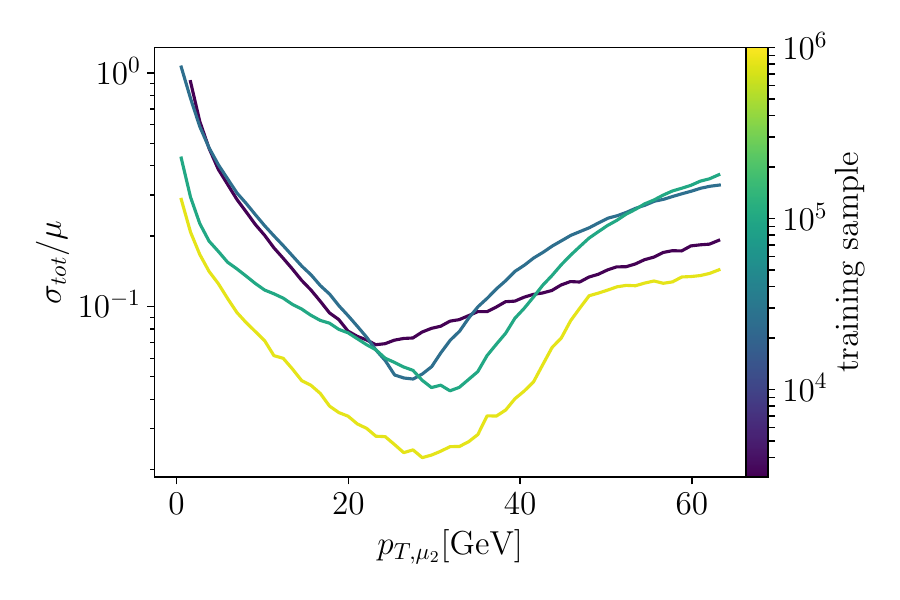}
    \includegraphics[width=0.495\textwidth,page= 3]{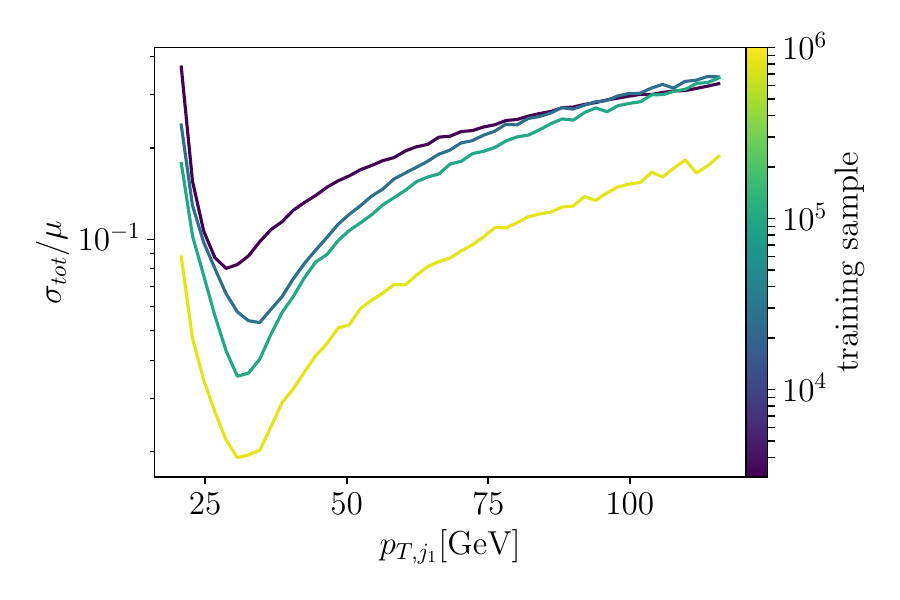}
    \caption{Relative uncertainty from the BINN for the $Z+1$~jet
      sample, as a function of the size of the training sample.}
    \label{fig:binn_stat}
\end{figure}
%----------------------------------------------------------

From the above discussion it is clear that one way to test the BINN
uncertainties is to train the same network the same way, but on
training samples of different size. We start with one batch size, 3072
events, and increase the training sample to the maximum of 2.7M. For
$Z+1$~jet we show the relative uncertainty as a function of transverse
momenta, for instance, in Fig.~\ref{fig:binn_stat}. In both cases we
see that over most the distribution the uncertainty improves with the
training statistics. However, we also see that in the right tail of
the $p_{T,\mu_1}$ distribution the lowest-statistics trainings does
not estimate the uncertainty correctly. Again, this reflects the fact
that, if the network does not even have enough data to estimate the
density, it will not provide a reliable uncertainty estimate. For
$p_{T,j_1}$ this effect does not occur, even in the tails where the
network has to extrapolate eventually.

%%%%%%%%%%%%%%%%%%%%%%%%%%%%%%%%%%%%%%%%%%%%%%%%%%%%%%%%%%%%%%%%%%%%%%%%
\subsection{Conditional augmentations}
\label{sec:uncert_cond}

%----------------------------------------------------------
\begin{figure}[t]
  \centering
  \includegraphics[width=0.495\textwidth]{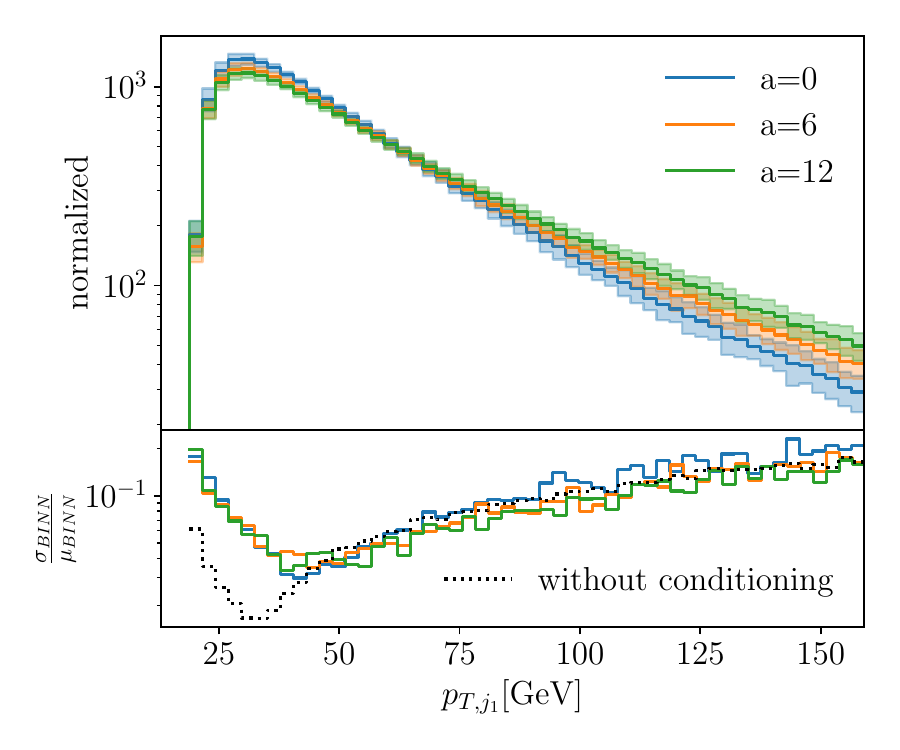}
   \includegraphics[width=0.495\textwidth]{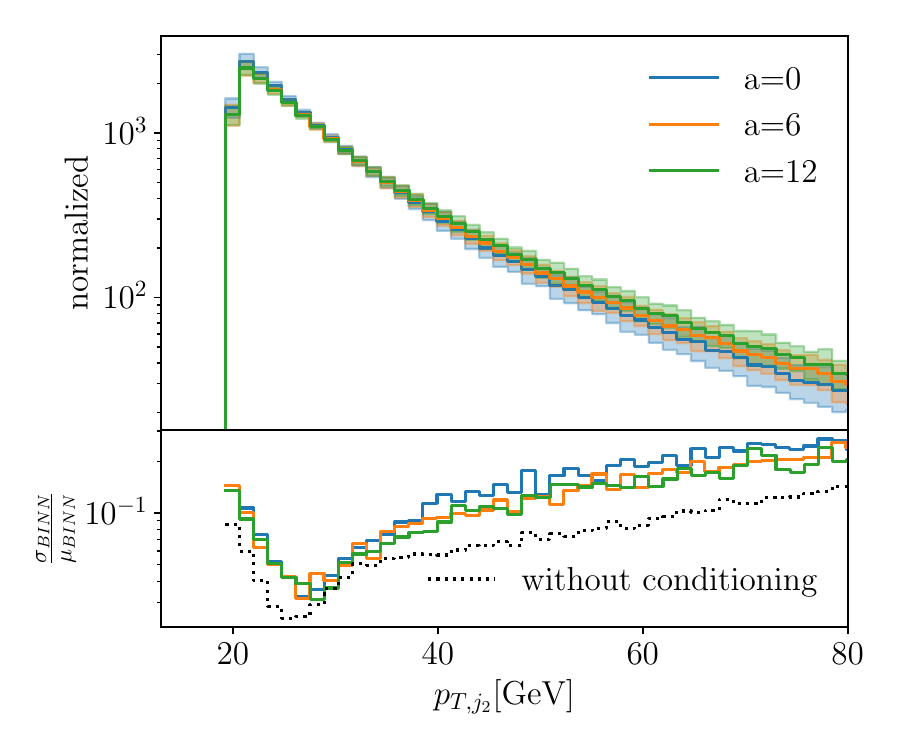}
  \caption{BINN densities for $Z+$~jets and conditional training with
    an enhanced-tail augmentation in $p_{T,_1}$, as defined in
    Eq.\eqref{eq:augment}.}
  \label{fig:binn_syst}
\end{figure}
%----------------------------------------------------------

As discussed above, Bayesian generative networks will not capture
typical systematic or theory uncertainties. Those uncertainties are
known, for instance as limitations to predict or reconstruct objects
in certain phase space regions, but unlike for regression or
classification networks we cannot augment the training date to
account for them. The reason is that generative networks extract the
underlying phase space density implicitly, so we cannot control what
augmented training data actually does to the network training.

For illustration purpose, let us introduce a toy theory uncertainty
proportional to the transverse momentum of a jet. This could incorporate
the limitation of an event generator, based on perturbative QCD, in
predicting tails of kinematic distributions inducing large logarithms.
In terms of a nuisance parameter $a$ such an uncertainty would shift
the unit weights of our training events to
\begin{align}
w = 1 + a \;  \left( \frac{p_{T, j_1} -15~\gev}{100~\gev} \right)^2 \; ,
\label{eq:augment}
\end{align}
where the transverse momentum is given in GeV, we account for a
threshold at 15~GeV, and we choose a quadratic scaling to enhance the
effects of this shift in the tails.

Instead ot just augmenting the training data, we train the network
conditionally on this nuisance parameter and then sample the nuisance
parameter for the trained network, to reproduce the systematic or
theory uncertainty now encoded in the network.  This means we then our
Bayesian INN conditionally on values $a = 0~...~30$ in steps of one.
For the event generation incorporating the theory uncertainty we can
sample kinematic distributions for different $a$-values. In
Fig.~\ref{fig:binn_syst} we show generated distributions for different
values of $a$. To model the conditional parameter similar to phase
space and allow for an uncertainty on the conditional nuisance
parameter, we sample $a$ with a Gaussian around its central value and
a standard deviation of $\min (a/10,0.1)$. The two panels show the
modified $p_{T,j_1}$-distribution and its impact on $p_{T,j_2}$
through correlations. As expected, the effects are similar, but the
multi-particle recoil washes out the effects on $p_{T,j_2}$. In the
upper panels we compare the effect of the theory uncertainty $a =
0~...~12$ to the statistical training uncertainty given by the
BINN. We see that our method traces the additional theory or
systematic uncertainty, and allows us to reliably estimate its
sub-leading nature for $p_{T,j_2}$. While we show ranges of $a$,
corresponding to the typical flat likelihood used for theory
uncertainties, we could obviously sample the different $a$-values
during event generation. In the lower panels we show the relative BINN
uncertainties, to ensure that the training for the different
$a$-values is stable. For $p_{T,j_1}$ the data augmentation has a
slight chilling effect on the high-precision training around the
maximum of the distribution. In the statistically limited tails
towards larger $p_T$ the BINN training without and with augmentations
behaves the same.  Looking at the recoil correlation, the BINN reports
a slightly larger uncertainty for the augmented training, correctly
reflecting the fact that the network now has to learn an additional
source of correlations. At least for the range of shown $a$-values
this BINN uncertainty is independent of the size of the augmentation.

%%%%%%%%%%%%%%%%%%%%%%%%%%%%%%%%%%%%%%%%%%%%%%%%%%%%%%%%%%%%%%%%%%%%%%%%
\subsection{Discriminator for consistency}
\label{sec:uncert_disc}

%----------------------------------------------------------
\begin{figure}[t]
  \centering
  \includegraphics[width=0.495\textwidth, page=3]{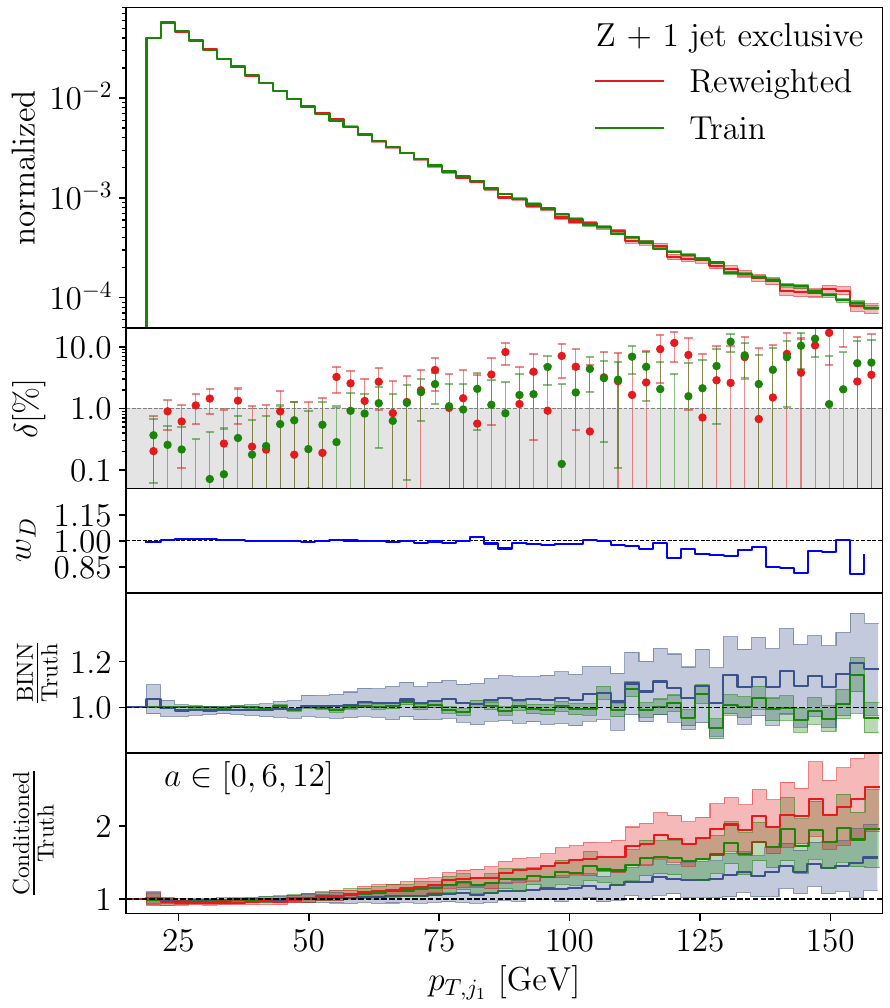}
  \caption{Illustration of uncertainty-controlled DiscFlow simulation. We
    show the reweighted $p_{T,j_1}$-distribution for the inclusive
    $Z$+jets sample, combined with the discriminator $D$, the BINN
    uncertainty, and the sampled systematic uncertainty defined
    through the data augmentation of Eq.\eqref{eq:augment}.}
  \label{fig:final}
\end{figure}
%----------------------------------------------------------

After introducing two ways of tracing specific uncertainties for
generative networks and controlling their precision, we come back to
the joint DiscFlow generator--discriminator training. In complete
analogy to, for instance, higher-order perturbative corrections, we
can use the jointly trained discriminator to improve the network
precision and at the same time guide us to significant differences
between training data and generated data. Because the discriminator is
a simpler network than the INN-generator, it is well suited to search
for deviations which the BINN misses in its density and uncertainty
maps.

In Fig.~\ref{fig:final} we illustrate the different aspects of our
uncertainty-controlled precision generator. First, we see that the INN
generator indeed learns and reproduces the phase space density at the
level of the training statistics. In the remaining panels we show
three ways to control possible uncertainty, using the discriminator, a
BINN, and a BINN combined with augmented training data. Each aspect is
described in detail in this paper:
\begin{itemize}
\item[$\cdot$] joint discriminator--generator training (DiscFlow) for precision generation --- Fig.~\ref{fig:discflow};
\item[$\cdot$] discriminator to control inconsistencies between training and generated events --- Fig.~\ref{fig:discflow_re};
\item[$\cdot$] BINN to track uncertainty on the learned phase space density --- Fig.~\ref{fig:binn};
\item[$\cdot$] conditional augmentation for systematic or theory
  uncertainties --- Fig.~\ref{fig:binn_syst}.
\end{itemize}
%

%%%%%%%%%%%%%%%%%%%%%%%%%%%%%%%%%%%%%%%%%%%%%%%%%%%%%%%%%%%%%%%%%%%%%%%%
\section{Outlook}
\label{sec:cln}

A crucial step in establishing generative networks as event generation
tools for the LHC is the required precision in estimating the phase
space density and full control of uncertainties in generated samples.

In the first part of this paper, we have shown how INN-generators can
be trained on $Z$+jets events with a variable number of particles in
the final state, to reproduce the true phase space density at the
percent level, almost on par with the statistical uncertainty of the
training sample. If we are willing to work with weighted events, with
event weights of order one, we can either use a magic variable
transformation or an additional discriminator network to achieve high
precision all over phase space. Alternatively, we can train the
discriminator jointly with the generator and use our novel DiscFlow
architecture to provide unweighted events with high precision
(Fig.~\ref{fig:discflow}). This joint training does not involve a Nash
equilibrium and is especially stable. Any information that the
discriminator has not transferred to the generator training can
eventually be included through reweighting, giving our NN-event
generator high precision combined with a high level of control
(Fig.~\ref{fig:discflow_re}).

In the second part of this paper we have established three methods to
control the precision INN-generator and its uncertainties. First, for
unsupervised generative training we can use a Bayesian INN to estimate
uncertainties from limited training statistics or sub-optimal network
training (Fig.~\ref{fig:binn}). Second, we can augment the training
data conditionally on a nuisance parameter and sample this parameter
to account for systematic or theory uncertainties including the full
phase space correlations (Fig.~\ref{fig:binn_syst}). A reliable
estimate of the different uncertainties allows us to compare the
numerical impact of the different uncertainties. Finally, we can use
the jointly trained discriminator to identify phase space regions
where the BINN lacks the necessary precision in its density and
uncertainty maps over phase space.

All these aspects of our uncertainty-controlled precision generator
are illustrated in Fig.~\ref{fig:final}.  With this level of precision
and control, INN-generators should be ready to be used as extremely
efficient tools to generate LHC events. More generally, our study
shows that generative INNs working on reconstructed objects can be
used as reliable precision tools for a range of forward and inverse
inference approaches as well as dedicated detector simulations.

%%%%%%%%%%%%%%%%%%%%%%%%%%%%%%%%%%%%%%%%%%%%%%%%%%%%%%%%%%%%%%%%%%%%%%
\begin{center} \textbf{Acknowledgments} \end{center}

We would like to thank Ben Nachman and Jan Pawlowski for very helpful
discussions on the DiskFlow loss function.  In addition, we would like
to thank Michel Luchmann and Manuel Haußmann for help with Bayesian
networks and Luca Mantani and Ramon Winterhalder for their work on an
earlier incarnation of this project. We are also very grateful to Ulli
K\"othe and Lynton Ardizzone for their expert advice on many aspects
of this paper.  The research of AB and TP is supported by the Deutsche
Forschungsgemeinschaft (DFG, German Research Foundation) under grant
396021762 – TRR 257 Particle Physics Phenomenology after the Higgs
Discovery.  TH is supported by the DFG Research Training Group
GK-1940, Particle Physics Beyond the Standard Model.  The authors
acknowledge support by the state of Baden-Württemberg through bwHPC
and the German Research Foundation (DFG) through grant no INST
39/963-1 FUGG (bwForCluster NEMO). This work was supported by the
Deutsche Forschungsgemeinschaft (DFG, German Research Foundation)
under Germany's Excellence Strategy EXC 2181/1 - 390900948 (the
Heidelberg STRUCTURES Excellence Cluster).

\bibliography{literature}

\end{document}